%% file: 100316D_SNbh.tex
\documentclass[structabstract]{aa}
\usepackage{graphicx}
\usepackage{natbib}
\bibpunct{(}{)}{;}{a}{}{,} % to follow the A&A style
\usepackage{hyperref}
\usepackage{setspace}  % singlespace-onhalfspace-doublespace
\usepackage{savesym}
\usepackage{amsmath}  % for the align environment
\savesymbol{iint}
\usepackage{txfonts}
\restoresymbol{TXF}{iint}
\usepackage[normalem]{ulem}
\usepackage[small,bf,nooneline,labelsep=period]{caption}
\newcommand{\rang}{\mbox{\,--\,}}

\newcommand{\kps}{km s$^{-1}$}
\newcommand{\kpsM}{km s$^{-1}$ Mpc$^{-1}$}
\newcommand{\eps}{erg s$^{-1}$}
\newcommand{\nodata}{~$\cdots$~}
\newcommand{\griz}{$g\,'r\,'i\,'z\,'\!$}
\newcommand{\gri}{$g\,'r\,'i\,'\!$}
\newcommand{\gr}{$g\,'r\,'\!$}
\newcommand{\gz}{$g\,'z\,'\!$}
\newcommand{\ri}{$r\,'i\,'\!$}

\newcommand{\grizJ}{$g\,'r\,'i\,'z\,'J$}
\newcommand{\grizJH}{$g\,'r\,'i\,'z\,'JH$}

\newcommand{\JHK}{$JHK_{\rm s}$}
\newcommand{\HK}{$HK_{\rm s}$}
\newcommand{\K}{\ensuremath{K_{\rm s}}}
\newcommand{\EBVhost}{\ensuremath{E(B-V\,)_{\rm host}}}
\newcommand{\EBVGal}{\ensuremath{E(B-V\,)_{\rm Gal}}}

\newcommand{\NHhost}{\ensuremath{N_{\rm\,H,host}}}
\newcommand{\NHGal}{\ensuremath{N_{\rm\,H,Gal}}}
\begin{document}

%% more of my own macros:
\newcommand{\Ahost}[1]{\ensuremath{A_{#1{\rm,host}}}}
\newcommand{\prima}[1]{\ensuremath{#1\,'\!}}
\newcommand{\gmr}{\ensuremath{\prima{g}-\prima{r}}}
\newcommand{\gmz}{\ensuremath{\prima{g}-\prima{z}}}
\newcommand{\rmi}{\ensuremath{\prima{r}-\prima{i}}}
\newcommand{\rmz}{\ensuremath{\prima{r}-\prima{z}}}
\newcommand{\rmJ}{\ensuremath{\prima{r}-J}}
\newcommand{\imJ}{\ensuremath{\prima{i}-J}}
\newcommand{\JmH}{\ensuremath{J-\!H}}

\title{The fast evolution of SN 2010bh associated with XRF 100316D}
%\subtitle{}

\author{F.~Olivares~E.\inst{1}\fnmsep\thanks{\email{foe@mpe.mpg.de}},
  J.~Greiner\inst{1}, P.~Schady\inst{1}, A.~Rau\inst{1},
  S.~Klose\inst{2}, T.~Kr\"uhler\inst{1,3,4},
  P.~M.~J.~Afonso\inst{1}\fnmsep\thanks{Present address: American
    River College, Physics \& Astronomy Dpt., 4700 College Oak Drive,
    Sacramento, CA 95841, USA}, A.~C.~Updike\inst{5,6},
  M.~Nardini\inst{1}\fnmsep\thanks{Present address: Universit\`a degli
    studi di Milano-Bicocca, Piazza a della Scienza 3, 20126, Milano,
    Italy}, R.~Filgas\inst{1}, A.~Nicuesa~Guelbenzu\inst{2},
  C.~Clemens\inst{1}, J.~Elliott\inst{1}, D.~A.~Kann\inst{2},
  A.~Rossi\inst{2}, \and V.~Sudilovsky\inst{1}}

\institute{Max-Planck-Institute f\"ur extraterrestrische Physik,
  Giessenbachstra\ss{}e 1, 85748 Garching, Germany
  \and
  Th\"uringer Landessternwarte Tautenburg, Sternwarte 5, 07778
  Tautenburg, Germany
  \and 
  Universe Cluster, Technische Universit\"at M\"unchen,
  Boltzmannstra\ss{}e 2, 85748 Garching, Germany
  \and
  Dark Cosmology Centre, Niels Bohr Institute, University of
  Copenhagen, Juliane Maries Vej 30, 2100 Copenhagen, Denmark
  \and
  CRESST and the Observational Cosmology Laboratory, NASA/GSFC,
  Greenbelt, MD 20771, USA
  \and
  Department of Astronomy, University of Maryland, College Park, MD
  20742, USA}
 
\date{Received August 23, 2011; accepted October 31, 2011}

  \input{abst}

  \titlerunning{SN 2010bh alias XRF 100316D}
  \authorrunning{Olivares E. et al.}

  \maketitle
%  \tableofcontents

%%%%%%%%%%%
\input{c1_intro}
\input{c2_obs}
\input{c3_bbsed}
\input{c4_multi}
\input{c5_bolo}
\input{c6_concl}
\input{zAck}

%\bibliographystyle{aa} % style aa.bst
%\bibliography{zRef} % your references Yourfile.bib

\input{100316D_SNbh.bbl}
\Online
\input{zA1_data}

\end{document}

%% file: abst.tex
% \abstract{}{}{}{}{} 
% 5 {} token are mandatory
 
  \abstract
  % context heading (optional)
  % {} leave it empty if necessary
  { The first observational evidence of a connection between
    supernovae (SNe) and gamma-ray bursts (GRBs) was found about a
    decade ago. Since then, only half a dozen spectroscopically
    confirmed associations have been discovered and XRF 100316D
    associated with the type-Ic SN 2010bh is among the latest.}
  % aims heading (mandatory)
  { We constrain the progenitor radius, the host-galaxy extinction,
    and the physical parameters of the explosion of XRF 100316D/SN
    2010bh at $z=0.059$. We study the SN brightness and colours in the
    context of GRB-SNe.}
  % methods heading (mandatory)
  { We began observations with the Gamma-Ray burst Optical and
    Near-infrared Detector (GROND) 12 hours after the GRB trigger and
    continued until 80 days after the burst. GROND provided excellent
    photometric data in six filter bands covering a wavelength range
    from approximately 350 to 1800 nm, significantly expanding the
    pre-existing data set for this event.  Combining GROND
    and \emph{Swift} data, the early broad-band spectral energy
    distribution (SED) is modelled with a blackbody and afterglow
    component attenuated by dust and gas absorption.  The temperature
    and radius evolution of the thermal component are analysed and
    combined with earlier measurements available from the literature.
    Templates of SN 1998bw are fitted to the SN itself to directly
    compare the light-curve properties.  Finally, a two-component
    parametrised model is fitted to the quasi-bolometric light curve,
    which delivers physical parameters of the explosion.}
  % results heading (mandatory)
  { The best-fit models to the broad-band SEDs imply moderate
    reddening along the line of sight through the host galaxy
    ($\Ahost{V}=1.2\pm0.1$ mag).  Furthermore, the parameters of the
    blackbody component reveal a cooling envelope at an apparent
    initial radius of $7\times10^{11}$ cm, which is compatible with a
    dense wind surrounding a Wolf-Rayet star. A multicolour comparison
    shows that SN 2010bh is 60\rang70\% as bright as SN
    1998bw. Reaching maximum brightness at 8\rang9 days after the
    burst in the blue bands, SN 2010bh proves to be the most rapidly
    evolving GRB-SNe to date. Modelling of the quasi-bolometric light
    curve yields $M_{\rm Ni}=0.21\pm0.03 M_\odot$ and $M_{\rm
    ej}=2.6\pm0.2 M_\odot$, typical of values within the GRB-SN
    population. The kinetic energy is $E_{\rm
    k}=(2.4\pm0.7)\times10^{52}$ erg, which is making this SN the
    second most energetic GRB-SN after SN 1998bw.}
  % conclusions heading (optional), leave it empty if necessary
  { This supernova has one of the earliest peaks ever recorded and
    thereafter fades more rapidly than other GRB-SNe, hypernovae, or
    typical type-Ic SNe. This could be explained by a thin envelope
    expanding at very high velocities, which is therefore unable to
    retain the $\gamma$-rays that would prolong the duration of the SN
    event.}


%% file: c1_intro.tex
\section{Introduction}

%% FACTS BEFOREHAND %%
The core collapse of massive stars is thought to give rise to
supernovae (SNe) and long-duration gamma-ray bursts \citep[GRBs,
durations larger than 2~s;][]{Kouveliotou93}. The first clue to the
connection between these events is the similarity in their
kinetic-energy scale (see \citealt{WooBlo06} for a review). Even
before the discoveries of the late 90's, a few authors discussed the
possible association between high-energy outbursts and SNe
\citep{Colgate68,Paczynski86}. In theory, the collapsing core of a
very massive star, whose envelope has been blown away by its own
stellar winds, i.e., a Wolf-Rayet (WR) star, can lead to the formation
of a relativistic jet that will produce high-energy emission
\citep{Woosley93,WooMac99} in the form of a GRB or an X-ray flash
\citep[XRF, softer events thought to be mainly produced by the same
mechanism as GRBs;][]{Heise01,Kippen04,Sakamoto05,HeiZan01}.  When the
jet collides with the circumstellar material, it produces a
multi-wavelength afterglow \citep[for a review,
see][]{ZhaMes04}. Theoretically, the energy of the core collapse
should also be capable of expanding the remaining envelope. However,
it is unclear how this can happen, or even if there is always enough
energy for a SN explosion. The kind of SN thought to be associated
with GRBs are those called stripped-envelope (SE) SNe, whose hydrogen
envelope have mostly been removed. Events such as these are unusually
energetic stellar explosions and sometimes labelled ``hypernovae'' in
the literature \citep[e.g.,][]{Paczynski98,Hansen99}.

%% FIRST EVIDENCE %%
To date, only type-Ic SNe have been related to GRBs. The first and
most representative case was that of SN 1998bw, which was found to be
associated with the soft GRB 980425
\citep[e.g.,][]{Galama98b,Kippen98b}. A couple of days after the
trigger, SN 1998bw was discovered \citep{Galama98a,Sadler98} inside
the $8'$ error circle of GRB 980425 \citep{Soffitta98} in the
underluminous late-type galaxy ESO184--G82
\citep[$z=0.0085$;][]{Tinney98}. Although initially controversial
\citep{Galama98b,Pian98}, the physical association between these
objects was supported on temporal and spatial grounds by the slowly
variable X-ray source at the position of the SN
\citep{Pian00,Kouveliotou04}. As a result of this connection, both the
GRB and SN research fields, which until then had both evolved more or
less independently, were revolutionised by a single event.

%% SPECTROSCOPY %%
Five years after SN 1998bw, the association between GRB 030329 and SN
2003dh came to light through a clear spectroscopic identification
\citep{Hjorth03,Kawabata03,Stanek03,Matheson03} and became the first
truly solid piece of evidence in favour of the GRB-SN connection.
Moreover, after 1998 there have been many other spectroscopic
associations, such as the cases of GRBs 021211 \citep[SN
2002lt;][]{DellaValle03}, 020903 \citep{Soderberg05,Bersier06}, 031203
\citep[SN 2003lw;][]{Malesani04}, 050525A \citep[SN
2005nc;][]{DellaValle06a}, 060218 \citep[SN 2006aj;][]{Pian06}, 081007
\citep[SN 2008hw;][]{DellaValle08}, 091127 \citep[SN
2009nz;][]{Berger11,Cobb10,Filgas11}, and 101219B \citep[SN
2010ma;][]{Sparre11} based on hypernova features in their spectra.

%% LATE-TIME BUMPS AND SAMPLES %%
Furthermore, late-time bumps in the light curves of GRB afterglows
have been interpreted as SN signals, e.g., GRBs 970228
\citep{Reichart99,Galama00a,ReiLamCas00}, 980326
\citep{CasGor99,Bloom99}, 011121 \citep{Bloom02,Greiner03}, 020405
\citep{Price03,Masetti03}, 040924 \citep{Soderberg06,Wiersema08},
041006 \citep{Stanek05,Soderberg06}, 050824 \citep{Sollerman07},
060729, and 090618 \citep[both in][]{Cano11b} to mention a few. These
bumps show consistency in terms of colour, timing, and brightness with
those expected for the GRB-SN population, but they are usually faint,
which hampers the spectroscopic identification. These re-brightenings
have been detected in GRB light curves out to redshifts of $\sim1$
\citep{Masetti05,DellaValle03,Bloom09,Tanvir10} owing to the
sensitivity of current ground-based telescopes dedicated to follow-up
observations. Sample studies of GRB-SNe (including bumps not
spectroscopically identified) have been carried out to determine the
luminosity distribution, the morphology of the light curves, and the
physical parameters of the explosion such as kinetic energy, ejected
mass, and $^{56}$Ni mass
\citep{ZehKloHar04,Ferrero06,Richardson09,Thoene11}. It has been
asserted that GRB-SNe are in general brighter than the local sample of
SE SNe, except for cases such as SN 2010ay \citep{Sanders11}. In
contrast to the claims of \citet{Stanek05}, no correlation has been
found between the brightness at maximum with the shape of the light
curve \citep{Ferrero06}.

%% SHOCK BREAKOUT %%
Additional information about the explosion can be obtained from its
early emission. One particular case is that of the soft XRF 060218
associated with SN 2006aj
\citep{Campana06,Pian06,Ferrero06,Cobb06,Modjaz06,Sollerman06,Mirabal06}.
This displayed an early X-ray and UV emission, which was interpreted
as thermal radiation produced by the shock breakout from the surface
of the progenitor
\citep{Colgate74,Falk78,KleChe78,MatMcK99,WaxMesCam07,NakSar10}. The
envelope is heated up and owing to expansion it starts to
adiabatically cool, which then shifts the emission to UV and optical
wavelengths. From the analysis of this a signal, it is possible to
constrain the afterglow component, derive both the temperature and
luminosity of the thermal component, and compute the apparent radius
of emission \citep[e.g.,][]{Thoene11}. Other examples include SNe
without detectable $\gamma$-ray emission that nevertheless exhibit
adiabatic cooling in the UV/optical and/or X-ray observations: SN
2008D \citep{Soderberg08,Malesani09,Modjaz09,Mazzali08}, SN 2008ax
\citep{Roming09}, SNLS-04D2dc \citep{Schawinski08}, and SN 2010aq
\citep{Gezari10}, to mention a few. As additional information, it has
been claimed that high-energy emission in GRB-SNe comes from
accelerated shock-breakout photons rather than highly relativistic
jets \citep{Wang07}.

%% NON-DETECTIONS %%
Whilst no SN signature is expected for short GRBs, which are thought
to be produced by the mergers of compact objects (for deep
non-detections, see the cases of GRB 050509B in \citealt{Hjorth05a}
and \citealt{Bloom06}, and GRB 050709 in \citealt{Fox05} and
\citealt{Hjorth05b}), there are a few supposedly long events where an
expected SN appearance was never detected. In the cases of GRBs 060505
\citep{Fynbo06,Ofek07} and 060614
\citep{Fynbo06,GalYam06,DellaValle06b}, there are very tight
constraints on the SN signature, which go down to 1\% as bright as SN
1998bw. The validity of the non-detections of SNe components for these
two GRBs as a classification tool is a point of controversy
\citep[e.g.,][]{Zhang09}. We refer to \citet{Kann11} for an exhaustive
discussion on GRBs 060505 and 060614, in the light of the optical
luminosity of their afterglows.

We analysed the optical and near-infrared (NIR) data of XRF 100316D
and its associated SN 2010bh. The paper is organised by summarising
the observations, data acquisition, reduction, and analysis in Sect.\
\ref{sObs}. The main results are treated separately as three different
sections. The modelling of the early broad-band spectral energy
distribution (SED) provides the progenitor radius and the host-galaxy
extinction and is presented in Sect.\ \ref{sSED}. The multi-wavelength
light and colour curves are analysed in Sect.\ \ref{sCurv} along with
comparisons with previous GRB-SN events. In Sect.\ \ref{sBol}, the
quasi-bolometric light curve is analysed and the physical parameters
of the explosion are derived. Finally we gather our conclusions in
Sect.\ \ref{sCon}.

%% file: c2_obs.tex
\section{Observations}\label{sObs}

\begin{figure*}
  \centering
  \includegraphics[bb=-77 270 687 524,width=\textwidth,clip]{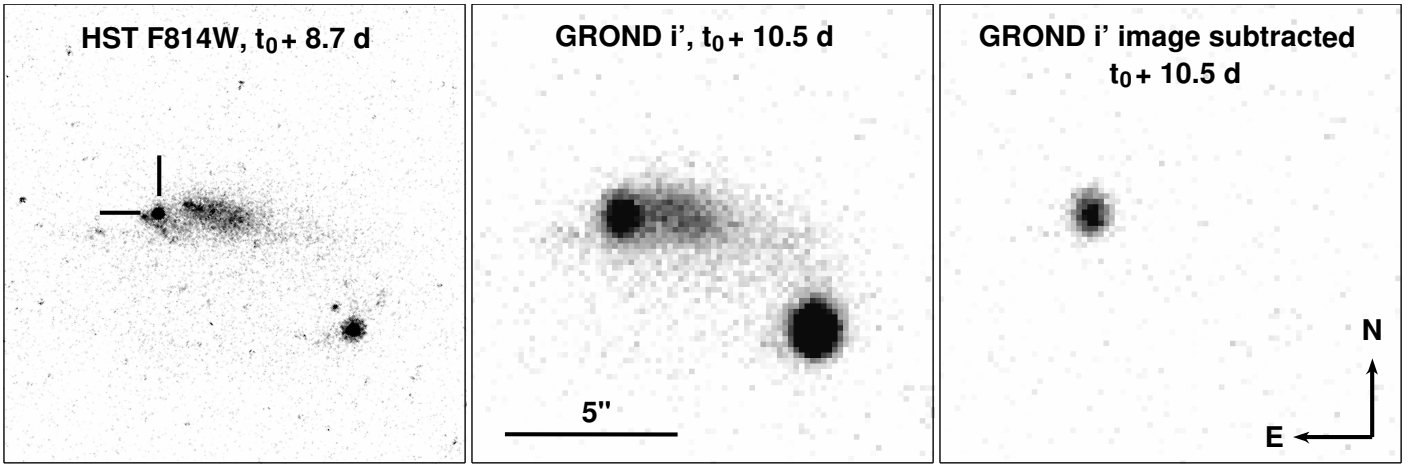}
  \caption{HST/WFC3 $F814W$, GROND \prima{i}-band, and GROND
    host-subtracted \prima{i}-band images of the GRB-SN field,
    respectively, from left to right. Images were taken at around
    maximum brightness. Each panel is approximately $13''\times13''$
    in size. The HST image shows significant galaxy structure near the
    explosion site (marked with two lines), which is blended with the
    object of interest in the GROND image shown in the middle
    panel. GROND images in the middle and right panels are shown using
    the same flux scale.}
  \label{fFC}
\end{figure*}

Gamma-ray emission from XRF 100316D triggered the Burst Alert
Telescope \citep[BAT;][]{Barthelmy05} on board the \emph{Swift}
satellite \citep{Gehrels04} on March 16, 2010, at $t_0=12$:44:50 UT
\citep{Stamatikos10}. It turned out to show a soft $\gamma$-ray
spectrum \citep{Sakamoto10} and a duration of at least 1300~s, one of
the longest ever measured \citep{Fan11,Starling11}. About 15 h
thereafter, a spectroscopic redshift of 0.059 was published for the
host galaxy \citep{Vergani10a,Vergani10b}. Observations of the
Gamma-Ray burst Optical and Near-infrared Detector
\citep[GROND;][]{Greiner07,Greiner08} confirmed that the new source
became evident about 16 h after the burst \citep{Afonso10}. The rising
of the supernova was verified photometrically by \citet{Wiersema10}
only three days after the trigger. The spectroscopic confirmation of
\citet{Chornock10a} came approximately six days after the burst, which
was confirmed two days later by \citet{Bufano10a}. The SN was
officially named SN 2010dh eight days after the trigger
\citep{Bufano10b,Chornock10b}. On March 26, 2010, additional GROND
observations of SN 2010bh were reported along with results at the
first attempts of host-galaxy subtraction \citep{Rau10}.

\subsection{GROND}

The multi-channel imager GROND \citep{Greiner07,Greiner08}, mounted at
the MPG/ESO 2.2m telescope on La Silla, Chile, started observations of
XRF 100316D 11.7 h after the trigger, simultaneously in \griz\JHK,
with an average seeing of 1\,\farcs1, as soon as the astronomical
night began.

The GROND data were reduced using standard {\tt
  pyraf/IRAF}\footnote{IRAF, the Image Reduction and Analysis
  Facility, is distributed by the National Optical Astronomy
  Observatory, which is operated by the Association of Universities
  for Research in Astronomy (AURA), Inc., under cooperative agreement
  with the National Science Foundation (NSF); see
  \href{http://iraf.noao.edu}{http://iraf.noao.edu}.} tasks
\citep{Tody93}, similar to the procedure described in detail by
\citet{Kruehler08}. A general point-spread function (PSF) model was
constructed using bright field stars, from which the full width at
half maximum (FWHM) was derived for each image. Aperture photometry
was performed for science and calibration objects using an aperture
size equal to the FWHM. The SDSS field at coordinates R.A.(J2000)
$=06^\mathrm{h}59^\mathrm{m}33^\mathrm{s}.6$, Dec.(J2000)
$=-17^\circ27'00''$ was observed during photometric conditions to
calibrate our images by performing relative photometry. A total of six
stars in the field of XRF 100316D were employed for this purpose (see
Table \ref{tStd} of the online material). The same set of stars was
used to calibrate \JHK\ against the 2MASS catalogue. Calibration
uncertainties vary in the range 0.002\rang0.020 mag for \griz\ and
0.02\rang0.12 mag for \JHK, which together with catalogue systematics
are added in quadrature to the statistical error.

A deep host-galaxy observation was carried out on November 5, 2010.
This observation resulted in images with mean seeing of 0\,\farcs74
and was used as a reference image for the subtraction of the host
contribution from the early epochs. To develop a notion of the host
contamination, we refer to Fig.\ \ref{fFC}, where the position of the
transient relative to the host galaxy is shown along with an $F814W$
image of the Wield Field Camera 3 (WFC3) on board the Hubble Space
Telescope (HST) as reference. In addition, the coordinates of the
centre of the host galaxy, catalogue name Anon J071031--5615, were
obtained from the last GROND observation: R.A.(J2000)
$=7^\mathrm{h}10^\mathrm{m}30^\mathrm{s}.37(\pm0^\mathrm{s}.07)$,
Dec.(J2000) $=-56^\circ15'20\,\farcs2(\pm0\,\farcs3)$. To align the
input and reference images, we use the {\tt WCSREMAP}
package\footnote{\href{http://www.astro.washington.edu/users/becker/wcsremap.html}{http://www.astro.washington.edu/users/becker/wcsremap.html}}.
For the main purpose of subtracting template from science images, the
{\tt HOTPANTS}
package\footnote{\href{http://www.astro.washington.edu/users/becker/hotpants.html}{http://www.astro.washington.edu/users/becker/hotpants.html}}
was employed. It matches the point spread function (PSF) and count
flux of both input images. It uses Gaussian functions to model the PSF
in sub-regions of the original image. Since reduced images are
essentially the combination of at least four \griz\ and 24 \JHK\
dithered individual exposures, three and five Gaussian functions of
different widths and degrees of freedom are employed for \griz\ and
\JHK, respectively. Point sources define the transformation that is
used to scale the template image to the flux of the science image. The
routine outputs a noise map of the resulting difference image, which
is employed to derive the uncertainties in the measured fluxes. The
package {\tt DAOPHOT/IRAF} was used to compute the magnitudes and
their corresponding errors. This procedure was executed for a total of
140 individual images in a total of 20 epochs (see Fig.\ \ref{fFC} for
an example of image subtraction in the \prima{i}\ band). The position
of the transient is R.A.(J2000)
$=7^\mathrm{h}10^\mathrm{m}30^\mathrm{s}.55(\pm0^\mathrm{s}.05)$,
Dec.(J2000) $=-56^\circ15'20\,\farcs0(\pm0\,\farcs2)$ in
host-subtracted optical images. The resulting photometry is tabulated
in Table \ref{tData} of the online material.

\subsection{\emph{Swift}/XRT and UVOT}

On board the \emph{Swift} satellite, the X-Ray Telescope
\citep[XRT;][]{Burrows05} and the UVOT \citep{Roming05} started
observations of XRF 100316D at $t_0+2.4$ min
\citep{Stamatikos10}. Whilst a bright X-ray source was detected inside
the BAT error circle, initially no afterglow candidate was found by
UVOT \citep{Oates10}.

However, in deeper images taken at $t_0+33$ ks in the $uvw1$ filter
and at $t_0+63$ ks in the $u$ band, we found evidence of emission in
excess of the host-galaxy contribution. To remove the contribution
from the host galaxy, we requested ToO observations in the $uvw1$ and
$u$ filters, which were taken at $t_0+3\times10^7$~s (347~d after the
burst), and amounted to a total exposure time of 1525 and 1369~s,
respectively. We measured a host galaxy contribution within the source
aperture of $25\pm2$ and $39\pm3\ \mu$Jy in the $uvw1$ and $u$ bands,
respectively. Subtracting this contribution from our earlier-time data
gave us a $3\sigma$ detection in 1894~s of $uvw1$- and in 4901~s of
$u$-band data taken at mid-times of $t_0+33$ ks and $t_0+63$ ks
respectively. In our analysis, we include only the $uvw1$ detection at
$t_0+33$ ks, since there were no GROND data contemporaneous with the
epoch of our $u$-band detection. All UVOT data were reduced following
the procedure described in \citet{Poole08}. To minimise the
contamination from the underlying host galaxy, the source flux was
measured within a circular source-extraction region of a 3\,\farcs5
radius. An aperture correction was then applied in order to remain
compatible with the UVOT effective area calibrations, which are based
on $5''$ aperture photometry \citep{Poole08}. The background was taken
from a source-free region with a $15"$ radius close to our source.

The relatively bright X-ray afterglow (30\rang40 cnt~s$^{-1}$ between
144 and 737~s after the trigger) faded considerably at the beginning
of the second XRT epoch \citep[$t_0+33$ ks; see][for a detailed
analysis]{Starling11}. In the subsequent analysis, we employed XRT
data at stages contemporaneous to GROND observations, specifically in
the interval from 33 to 508 ks after the burst. These data were
obtained from the public Swift archive and reduced in the standard
manner using the {\tt xrtpipeline} task from the HEAsoft package, with
response matrices from the most recent {\tt CALDB} release. All data
were obtained in photon counting mode and downloaded from the XRT
light curve repository \citep{Evans07,Evans09}. Spectra were grouped
using the {\tt grppha} task.

All the data discussed throughout the paper were corrected for the
Galactic-foreground extinction of $\EBVGal=0.117$ mag with $R_V=3.08$
\citep{SchFinDav98}. All uncertainties in the following analysis are
quoted at the $1\sigma$ confidence level.

%% file: c3_bbsed.tex
\section{Early broad-band SED}\label{sSED}

Using data between 33 and 54 ks (roughly from 11 to 15 h) after the
burst, we compiled two early broad-band SEDs of XRF 100316D. GROND
provides detections in \grizJ\ and upper limits in \HK, whilst the
count rate of contemporaneous \emph{Swift}/XRT observations is already
0.01 cnt s$^{-1}$ at $t_0+33$ ks and decaying. After combining data
from 33 to 508 ks after the trigger, XRT provides only three bins in
the X-ray energy range. As follows, XRT data are scaled to the GROND
first two epochs by using a decay index of $\alpha=-1.3\pm0.2$, which
is derived from the same XRT data over time. \emph{Swift}/UVOT
observed only in the $uvw1$ filter at these stages.

\subsection{Modelling scheme}\label{ssMod}

\begin{figure}
  \centering
  \includegraphics[bb=34 163 577 702,width=\linewidth,clip]{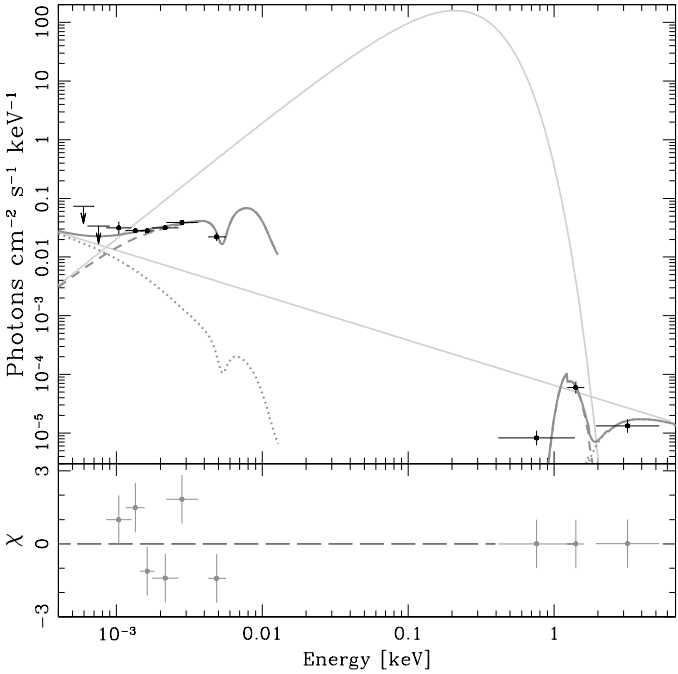}
  \caption{Broad-band SED at 50 ks after trigger. The observed data
    are represented by black filled circles. The \HK\ bands provide
    only $3\sigma$ upper limits shown as arrows. The thick grey line
    shows the best-fit model able to reproduce the data: extinguished
    power law plus blackbody components, which are shown individually
    with dotted and dashed grey lines, respectively. The thin
    continuous grey lines show the unextinguished versions of the
    power law corresponding to the afterglow and the thermal
    component. In the lower panel, the residuals of the best fit are
    plotted.}
  \label{fBB2}
\end{figure}

Early blue emission coming from the XRF position was detected by
GROND. For data acquired 42.5~ks after the burst, we measured a colour
$\gmr=-0.30\pm0.06$ mag, in contrast to the red afterglows that
usually follow a GRB (photon index $\Gamma$ in the range
1.2\rang2.5). Similar observations were made by \citet{Cano11a}, who
found that this emission is incompatible with synchrotron
radiation. The adiabatic cooling of the expanding atmosphere following
the shock breakout gives us a reasonable explanation of the observed
blue colours \citep[e.g.,][]{Cano11a}. In this scenario, the emission
from the shock breakout lasts only a few hours after the core collapse
and its SED resembles a blackbody at a high temperature of the order
of $10^6$~K \citep[or 0.1 keV equivalently; e.g.,][]{Campana06}. Thus,
we interpret the blue colours in our observations as the thermal
component associated with the cooling of the shock breakout, which is
similar to what has been claimed for XRF 060218/SN 2006aj
\citep{Campana06,WaxMesCam07} and other early-caught SNe (e.g.,
\citealt{Soderberg08,Modjaz09} on SN 2008D; \citealt{Roming09} on SN
2008ax).  In contrast to the idea of a thermal component producing the
observed emission, there are no significant contemporaneous detections
of UVOT in the $uvw1$ bandpass, which alludes to high reddening. On
the other hand, $\imJ=0.06\pm0.15$ mag measured at the same epoch
appears to have an underlying additive red component, which is
represented in this case by the GRB afterglow synchrotron emission
(see Fig.\ \ref{fBB2}).

\input{t1_mod}

To test this hypothesis, we modelled the SEDs at 42.5 and 50.0~ks (the
latter shown in Fig.\ \ref{fBB2}) using two additive components: (1) a
power law for the afterglow, and (2) an ideal blackbody for the
thermal component, respectively, of the forms
\begin{align}
  P_E(\Gamma)&=C_1\,E^{\,-\Gamma}\qquad\text{and}\\
  B_E(T_\mathrm{BB})&=\frac{C_2\,E^{\,2}\,dE}{(kT_\mathrm{BB}\,)^4(e^{\,E(1+z\,)/kT_\mathrm{BB}}-1)},
\end{align}
\noindent where $C_1$ and $C_2$ are normalisations, $E$ is the
spectral energy in units of keV, $\Gamma$ is the photon index, $k$ is
the Boltzmann constant, $T_\mathrm{BB}$ the intrinsic blackbody
temperature, and $z=0.059$ is the redshift \citep{Chornock10b}. The
model also accounts for host-galaxy extinction based on either Milky
Way (MW; $R_V=3.08$), Large Magellanic Cloud (LMC; $R_V=3.16$), or
Small Magellanic Cloud (SMC; $R_V=2.98$) extinction laws and soft
X-ray metal absorption. The Galactic metal absorption is fixed to be
$\NHGal=7.05\times10^{20}$ cm$^{-2}$ \citep{Kalberla05}. The model has
a total of six free parameters: $C_1$, $C_2$, $\Gamma$,
$T_\mathrm{BB}$, $\EBVhost$, and $\NHhost$. The luminosity of the
blackbody is computed as $L_\mathrm{BB}=8.0525\,C_2\,(1+z\,)D_{10}^{\
  2}$ \eps, where $D_{10}$ is the luminosity distance to the transient
in units of 10 kpc. A luminosity distance of $240\pm17$ Mpc to the
host galaxy of XRF 100316D is employed as computed by the NED
database\footnote{\href{http://nedwww.ipac.caltech.edu/}{http://nedwww.ipac.caltech.edu/}}
following the standard $\Lambda$CDM model, using the redshift measured
by \citet{Chornock10b}, a Hubble constant of 74.2 \kpsM\
\citep{Riess09}, and the model of the local velocity field by
\citet{Mould00}.

The different set of parameters are summarised in Table \ref{tMod}.
All best-fit parameters are consistent between the two epochs within
their statistical uncertainties between both epochs. The temperature
$T_\mathrm{BB}$ and luminosity $L_\mathrm{BB}$ of the thermal
component are in the range 78\rang81 eV and $4\rang7\times10^{47}$
\eps\ from Cols.\ 4 and 5 of Table \ref{tMod}, respectively. The
afterglow power-law photon index $\Gamma\approx1.8$ is shown in Col.\
7. The reduced $\chi\,^2$ (or $\chi_\mu\,^2\equiv\chi\,^2/\mu$, where
$\mu$ is the number of degrees of freedom) improves from 16.6 for the
model without extinction to 2.7 for the model extinguished by MW-like
dust. The best fit to the second SED epoch is shown in Fig.\
\ref{fBB2} with a thick grey line. The UV dust feature characteristic
of the MW extinction law gives the more precise results, although it
is poorly constrained at the bluer end. The X-ray tail of the
blackbody fits the two data bins at around 1 keV, whilst the power law
fits the only data point at $\approx3$ keV; both components contain
significant absorption. We note that the fitting of the X-ray data has
practically no residuals, i.e., the model over-predicts the data in
this energy range.

In addition to the luminosity and temperature, it is possible to
compute an apparent emission radius of the thermal component
($R_\mathrm{BB}$) from these two measured quantities by assuming
isotropic radiation from an ideal blackbody. For all our trial models,
the corresponding radii were calculated as $R_\mathrm{BB}=(4\pi\sigma
L_\mathrm{BB}^{\,-1}\,T_\mathrm{BB}^{\ 4})^{-1/2}$, where $\sigma$ is
the Stefan-Boltzmann constant (Col.\ 6 of Table \ref{tMod}). Our
best-fit parameters yield a radius of $3\rang4\times10^{13}$ cm, which
is two orders of magnitude larger than the typical sizes of the most
likely GRB progenitors \citep[WR stars;][]{CapGosVHu04}.

The amount of metal absorption (estimated based on an equivalent
hydrogen column density at solar metallicity) and dust extinction
required for a good fit are $\NHhost\approx4\times10^{22}$ cm$^{-2}$
and $\EBVhost=0.2\rang0.4$ mag, respectively. These two parameters are
fitted independently and with no assumption being made about the
environment gas-to-dust ratio. However, the available data is not
enough to allow us to distinguish among the different extinction laws
employed in the modelling procedure. Whilst the reddening is
consistent with values found in previous studies
\citep{Starling11,Cano11a}, the hydrogen column density differs
significantly from the results of \citet{Starling11}. The discrepancy
is due to the additional constraints provided by the optical/NIR data,
which were included to derive the parameters of the blackbody
contribution. With our data set it is possible to tie the thermal
component at both low and high energies, which provides a more
accurate value for the absorption by heavy elements because of the
greater constraint on the X-ray flux.

\subsection{Host-galaxy extinction}\label{ssExt}

\begin{figure}
  \centering
  \includegraphics[bb=30 83 765 414,width=\linewidth,clip]{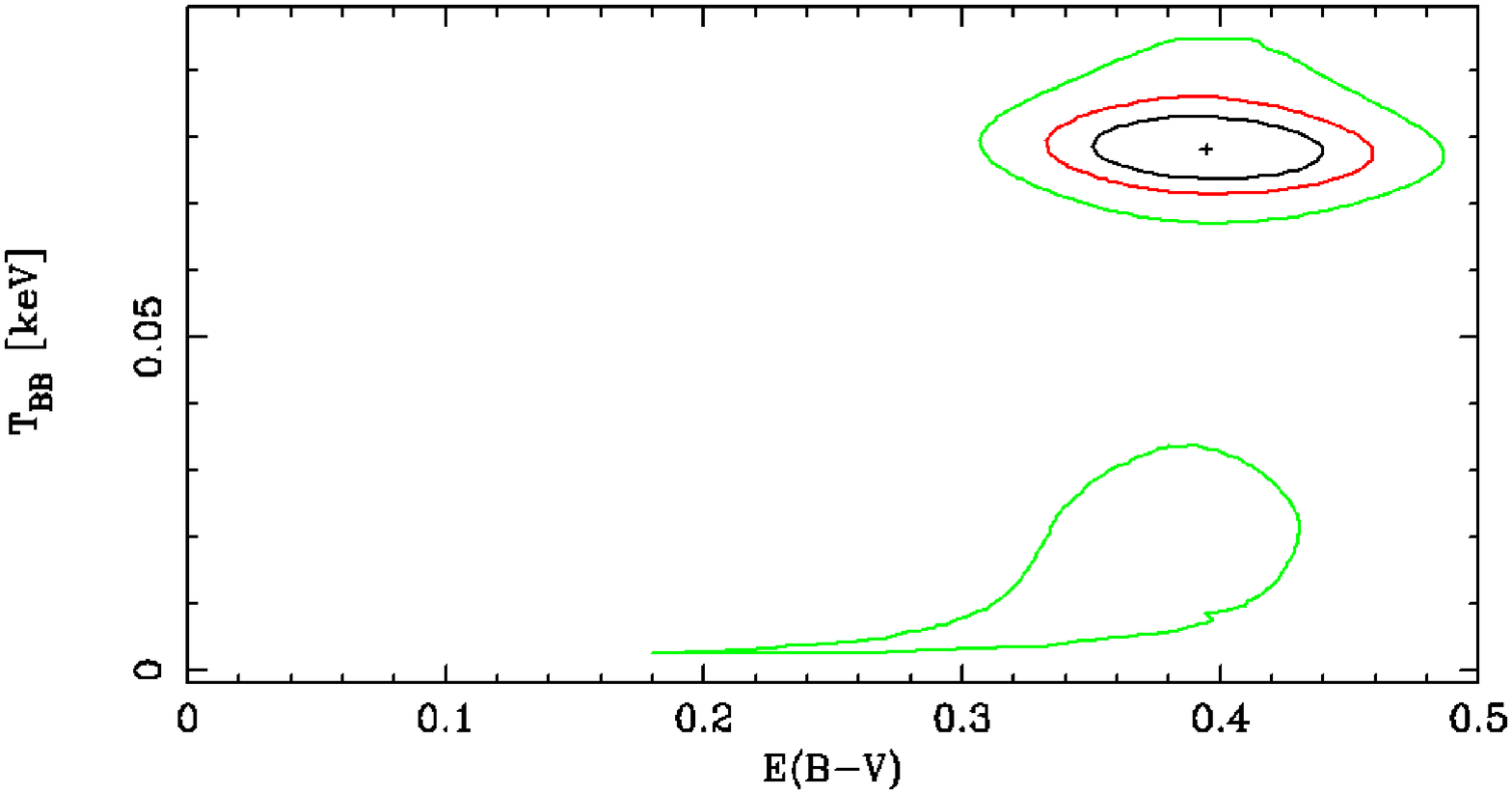}
  \includegraphics[bb=30 83 765 411,width=\linewidth,clip]{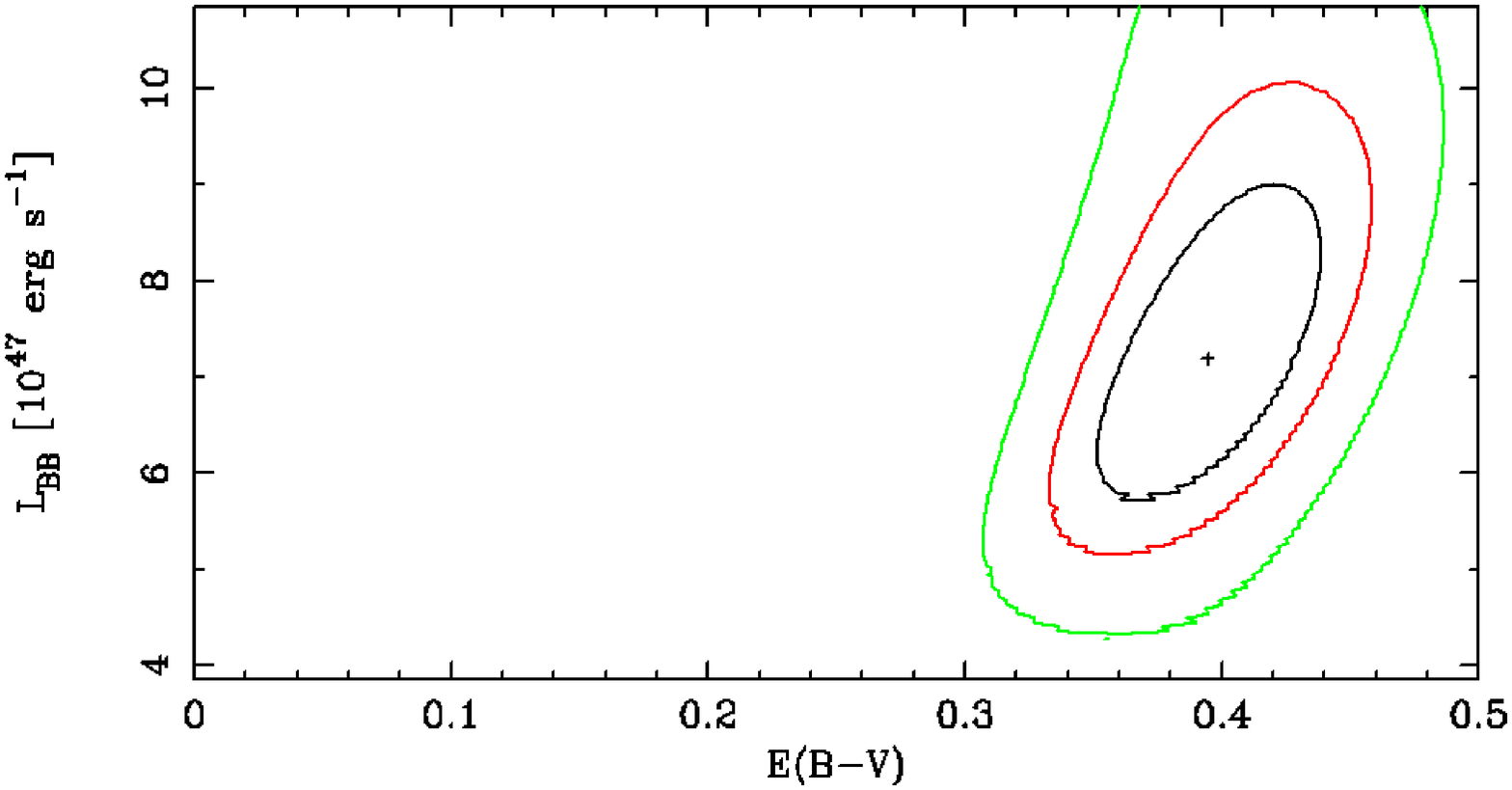}
  \includegraphics[bb=30 30 765 411,width=\linewidth,clip]{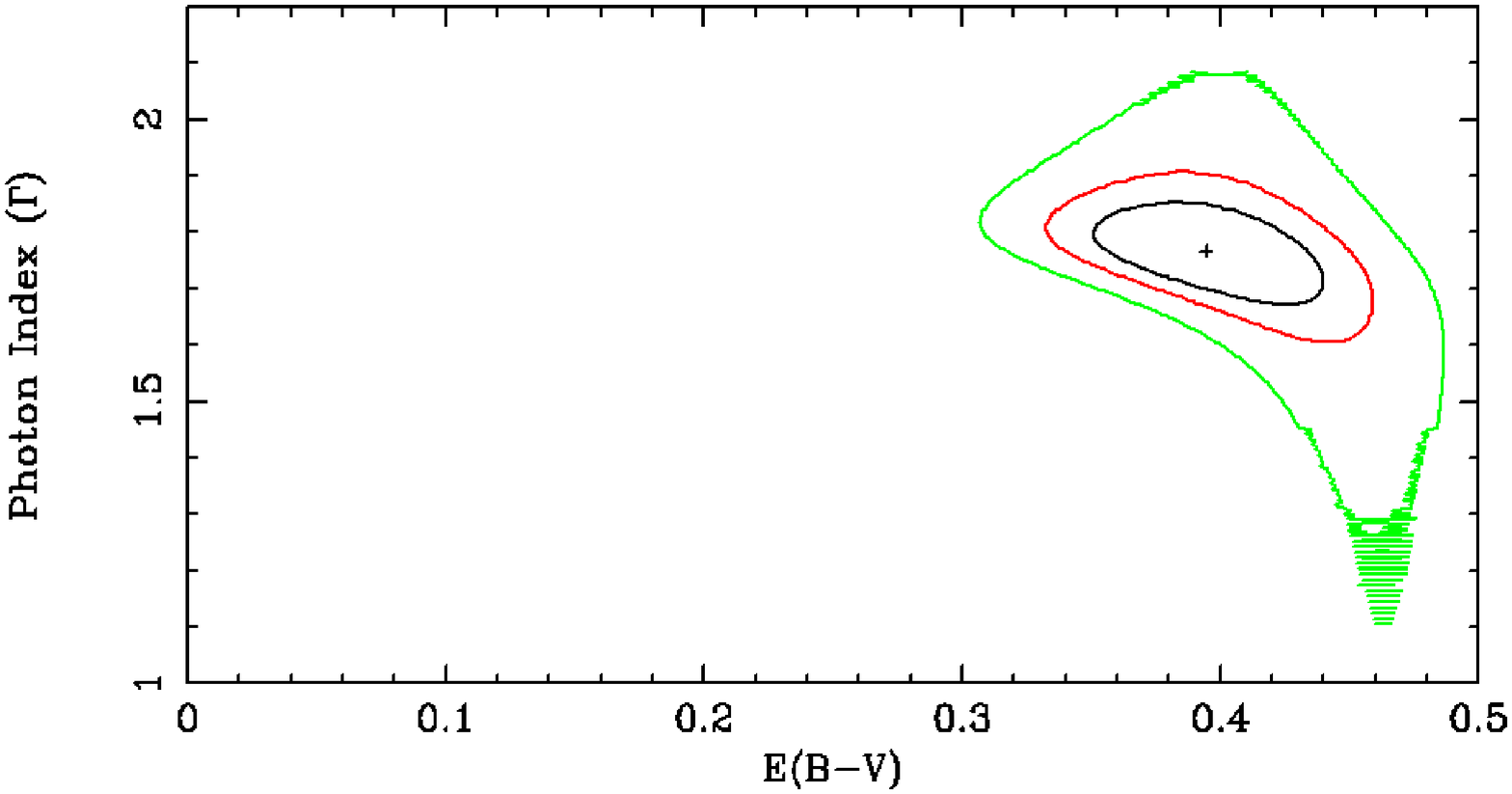}
  \caption{Contour plots for the best-fit parameters from the
    modelling of the second epoch. From the outer- to the innermost
    contours, the green, red, and black lines are 3, 2, and $1\sigma$
    contours, respectively. The tail at low $\Gamma$ in the bottom
    panel is caused by numerical inaccuracies.}
  \label{fCnt}
\end{figure}

We then attempted to evaluate the constraint on the extinction value
derived in the previous section. Contour plots of reddening against
blackbody temperature, luminosity, and spectral index from the best
fit to the data are shown in Fig.\ \ref{fCnt}, respectively. The
$1\sigma$ and $2\sigma$ contours show that the solution to our
best-fit relation is well-defined in all three panels. In the first
panel there is a second solution to our best-fit relation at the
$3\sigma$ level, which is at lower temperatures and similar reddening,
although its significance is rather low. In the middle panel, the
contours show a slight trend of proportionality between luminosity and
extinction, which is expected. At luminosities higher than
$1.1\times10^{48}$ \eps, there is no possible solution, because the
model becomes much brighter than the X-ray data and \NHhost\ cannot be
allowed to vary and compensate for the X-ray luminosity of the
model. Contours of the variations in the reddening and spectral index
are shown in the lower panel, where the extinction is again restricted
to high values.

%% ZERO EXTINCTION %%
Analyses of optical spectroscopy of the host galaxy have found that
the line-flux ratio of H$\alpha$ to H$\beta$ in the \ion{H}{ii} region
coincident with the SN position is consistent with zero extinction
\citep{Starling11,Levesque11}. Nevertheless, the $\Ahost{V}\sim0$
estimate from the spatially resolved spectroscopy of
\citet{Levesque11} probes a much larger region ($\approx1.3$ kpc$^2$)
than the one probed by our line of sight to the GRB-SNe. A
sufficiently high dust clumpiness could explain our high extinction
values along the line of sight and $\Ahost{V}\sim0$ when integrated
over a larger patch. The position at which the hydrogen lines are
formed, i.e., the \ion{H}{ii} region, might be located in front of the
explosion site and not probe the same line of sight through the host
galaxy.

%% BROKEN POWER LAW %%
Alternatively a broken power-law model was fitted to the data in
Fig.\ \ref{fBB2} for which $\chi_\mu\,^2=3.0$, which is slightly
larger than that of the model that consists of a blackbody plus power
law. The broken power law provided a closer fit when we assume that
there is no host-galaxy extinction, significantly different from our
results for the blackbody plus power-law model. However, the
low-energy spectral slope of $\beta=+0.5$ is incompatible with
synchrotron radiation \citep[$\,\beta_{\rm max}=+1/3$;
  e.g.,][]{SarPirNar98}. Furthermore, the spectral break lies between
the \prima{g}\ and the $uvw1$ bands at $\nu_{\rm
  break}=(8\pm3)\times10^5$ GHz, which is inconsistent with the
self-absorption feature usually observed at radio frequencies
\citep[$\nu_a\sim2\rang13$
  GHz;][]{Galama98c,Taylor98,Granot99,Galama00a}. Given also that
$\Ahost{V}>0.2$ mag is derived in Sect.\ \ref{ssMod} and preferred by
other authors (see next paragraph), the broken power-law model can be
discarded with confidence.

%% LARGER REDDENING VALUES %%
Additional evidence of large reddening along the line of sight inside
the host galaxy was found by \citet{Starling11}, $\EBVhost\approx0.9$
mag. The reddening was derived by fitting \emph{Swift}/BAT+XRT data
and a $u$-band $3\sigma$ upper limit provided by \emph{Swift}/UVOT
data in the interval from 638 to 737~s after the burst. We employed a
model that consists of a blackbody plus power law extinguished by SMC
dust. A similar method was used here and by \citet{Campana06}, who
determined $\EBVhost=0.20$ for SN 2006aj. Another attempt at
dereddening SN 2010bh was carried out by \citet{Cano11a}, who found
$\EBVhost=0.18\pm0.08$ mag by assuming that the colours of SE SNe are
all the same ten days after the $V$-band maximum brightness
\citep{Drout10}. Whilst this method is supported for the hydrogen
atmospheres of type-IIP SNe by a line of physical arguments
\citep[e.g.,][]{Olivares10}, it is entirely empirical for SE
SNe. Nevertheless, their reddening value is larger than zero with a
significance of $2.3\sigma$ and consistent with our calculations for
our second-epoch SED of $\EBVhost=0.39\pm0.03$ mag at the $2.5\sigma$
confidence level.

After correcting for $\EBVhost=0.39\pm0.03$ mag, we obtained
$uvw1=18.15\pm0.27$ and $u=19.70\pm0.45$ mag in the AB system at 33
and 63 ks after the burst, respectively. For SN 2006aj
\citep{Campana06}, these values were $uvw1=17.73\pm0.21$ and
$u=17.77\pm0.15$ mag at the same phase and redshift of SN 2010bh.  If
we had assumed that we should have seen a cooling envelope for SN
2010bh of comparable brightness and evolution as that shown by SN
2006aj, the host-galaxy extinction should have been higher than that
estimated by our method. Nevertheless, the uncertainties are large and
the comparison of the \emph{uvw1} measurements is consistent at the
$1.2\sigma$ confidence level. Although in the \emph{u} band the
uncertainty is even larger, there is no consistency with the
brightness of SN 2006aj at the $4\sigma$ confidence level and a higher
host-galaxy extinction ($\Ahost{V}\approx2.1$ mag) would be required
to reach the $1\sigma$ level of consistency.

%% CONCLUSION %%
In conclusion, we use the extinction value from the fit to the
second-epoch broad-band SED throughout the paper,
$\EBVhost=0.39\pm0.03$ mag for MW-like dust with $R_V=3.08$, given
that compared to the first epoch the statistical errors in the \grizJ\
photometry are smaller and additional $uvw1$ photometry is
available. Moreover, when fixing the host-galaxy reddening to
$\EBVhost=0.39$ mag, the quality of the fit to the first epoch is
still acceptable ($\,\chi_\mu\,^2=1.6$), whilst the fit to the second
epoch using $\EBVhost=0.21$ mag from the first-epoch modelling results
in $\chi_\mu\,^2=3.7$ and unphysical parameters. Hence, values of
host-galaxy extinction including the correction for redshift
(\emph{K}-correction based on the spectral model) for the GROND
filters and their corresponding statistical uncertainty are
$\Ahost{\prima{g}}=1.50\pm0.12$, $\Ahost{\prima{r}}=1.10\pm0.09$,
$\Ahost{\prima{i}}=0.80\pm0.08$, $\Ahost{\prima{z}}=0.60\pm0.06$,
$\Ahost{J}=0.39\pm0.04$, $\Ahost{H}=0.22\pm0.02$, and
$\Ahost{\K}=0.14\pm0.01$, all in units of magnitude.

\subsection{Evolution of the thermal component}

\begin{figure}
  \centering
  \includegraphics[bb=34 163 573 708,width=\linewidth,clip]{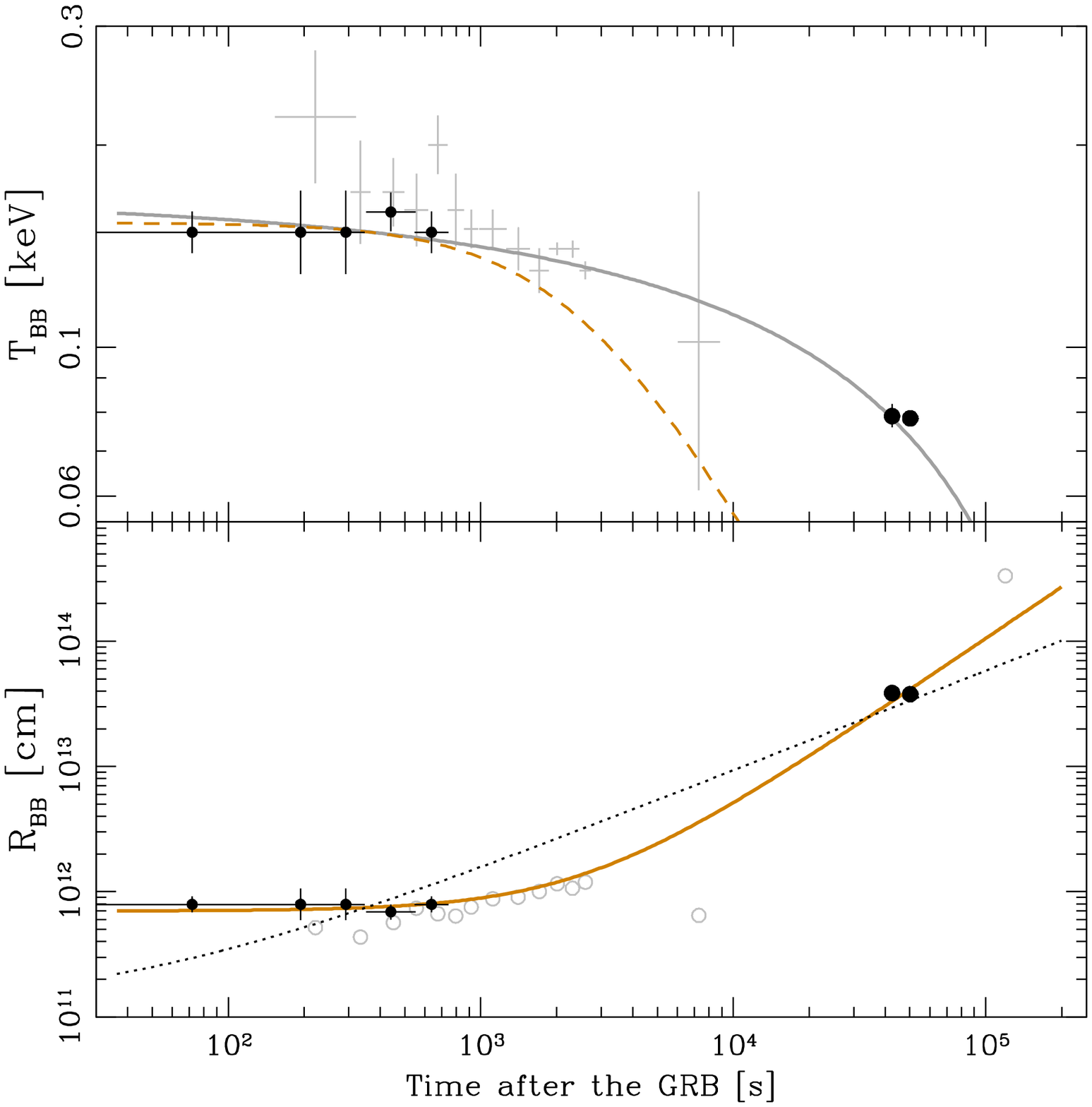}
  \caption{Temperature and radius evolution of the blackbody
    component. Black data points are those derived using combined
    BAT+XRT data \citep{Starling11} plus the temperature
    determinations using GROND data ($\approx46$ ks). The first data
    bin from \citet{Starling11} represents the interval from $-175$ to
    144~s, however, due to fitting and plotting purposes, here it
    corresponds to 0\rang144~s. Solid, dashed and dotted lines are
    different power-law models (see main text). Best fits to
    temperature and radius are shown in grey and brown solid lines,
    respectively. Grey crosses (upper panel) and open circles (lower
    panel)) are measurements of XRF 060218/SN 2006aj taken from
    \citet{Kaneko07} and \citet{Campana06}, respectively.}
  \label{fTR}
\end{figure}

Having determined temperature, luminosity, and radius for the thermal
component, it was then possible to study their
evolution. \citet{Starling11} analysed combined BAT+XRT data until
737~s after the trigger and derived blackbody temperatures, data that
we used in the following analysis. Temperature and radius over time
are shown in Fig.\ \ref{fTR}. The decrease in temperature and the
increase in radius are both trends that are consistent with the
cooling of the envelope due to expansion and, thus, with the explosive
scenario.

The temperature and radius of an expanding envelope that is cooling
adiabatically have been theoretically shown to evolve as power laws
\citep{WaxMesCam07,NakSar10}. We therefore fit power laws to the decay
and rise of temperature and radius to the combined data presented here
and in \citet{Starling11}. The expression
$T_\mathrm{BB\,}(t\,)=T_i-\kappa t^{\,\delta}$ was fitted to the
evolving temperature (solid line in the upper panel of Fig.\
\ref{fTR}), which gave a decay index of $\delta=0.3\pm0.2$ for an
initial temperature of $T_i=0.17\pm0.04$ keV with $\chi_\mu\,^2=0.8$,
where $\kappa$ is the normalisation of the power law.

A model of the form $R_\mathrm{BB}(t\,)=R_0+\upsilon t^{\,\gamma}$ was
employed to fit the radius measurements. In the case of linear
expansion, i.e., $\gamma=1$, the radius grows at a mean velocity of
$\upsilon\approx8,000$ \kps\ between BAT+XRT (until 737~s) and GROND
observations (at 42\rang54 ks after the burst).  When $\gamma$ was
allowed to vary, we obtained a growth index of $\gamma=1.4\pm0.3$ and
an initial radius of $R_0=(7.0\pm0.9)\times10^{11}$ cm with
$\chi_\mu\,^2=1.1$ (solid line in the lower panel of Fig.\ \ref{fTR}).
The resulting radius is slightly larger than the size of WR stars
\citep[$\sim10^{11}$ cm;][]{CapGosVHu04}, which are thought to be the
progenitors of long-duration gamma-ray bursts and type-Ic SNe
\citep{WooHegWea02}. Because of this, the initial emission radius of
the thermal component might indicate the position at which a
preexisting dense wind surrounding the progenitor becomes optically
thin \citep[e.g.,][]{Campana06,Soderberg08,BalLoe11}. From the
theoretical point of view, it is $\gamma=0.8$ \citep{WaxMesCam07},
although, the data do not favour this solution delivering
$\chi_\mu\,^2=8$ (dotted line in the lower panel of Fig.\ \ref{fTR}).

For an assumption of adiabatic cooling, the luminosity must be
constant and was fixed to $L_\mathrm{BB}=3.5\times10^{45}$ \eps, value
derived from early-time X-ray measurements
\citep{Starling11}. Assuming the best-fit model for the radius
evolution from the previous paragraph, we computed the evolution of
the adiabatic temperature as
$T_\mathrm{BB}=(4\pi\sigma\,L_\mathrm{BB}^{-1}\,R_\mathrm{BB}^{\,2}\,)^{-1/4}$,
which is shown by the dashed line in the upper panel of Fig.\
\ref{fTR}. This model is clearly inconsistent with our data set;
however, it indeed shows consistency with the XRF 060218/SN 2006aj
data set (grey crosses in the upper panel of Fig.\ \ref{fTR}; from
\citealt{Kaneko07}). Since the blackbody luminosity changes to
$4\rang6\times10^{47}$ \eps\ in our late-time measurements, the
assumption of a constant luminosity is invalid and inconsistent with
the late-time temperature determination.  This result implies that
either the cooling is not strictly adiabatic for which energy
injection from the inner core is needed, or $T_\mathrm{BB}$ and
$L_\mathrm{BB}$ are overestimated by our modelling procedure, or
underestimated by \citet{Starling11}.

%% file: t1_mod.tex
\begin{table*}
\begin{minipage}[t]{\textwidth}
  \caption{Fits to broad-band spectra constructed using GROND and
    \emph{Swift}/XRT data.}\label{tMod}
  \centering
%\begin{onehalfspace}
\begin{tabular}{r@{\,--\,}lcccccclr@{/}l}
 \hline\hline\noalign{\vspace{0.5\smallskipamount}}
 \multicolumn{2}{c}{Time Interval}      &Reddening  &\multicolumn{1}{c}{\EBVhost}  &$kT_\mathrm{BB}$  &$L_\mathrm{BB}$   &$R_\mathrm{BB}$ &$\Gamma$   &\NHhost             &$\chi^2$  &$\mu$ \vspace{0.5\smallskipamount}\\
 \multicolumn{2}{c}{[s] after the trigger} &Law        &\multicolumn{1}{c}{[mag]}      &[eV]         &[$10^{47}$ \eps]  &[$10^{12}$ cm]  &        &[$10^{22}$ cm$^{-2}$]  &\multicolumn{2}{c}{}  \vspace{0.5\smallskipamount}\\
 \hline\noalign{\smallskip}
 42182 &42879  &MW   &0.21$_{-0.21}^{+0.11}$  &$81_{-4}^{+6}$  &$4.1_{-2.5}^{+2.0}$  &27$\pm8$  &1.82$\pm0.05$  &4.2$\pm0.5$  &4.9  &4    \smallskip\\
 42182 &42879  &LMC  &0.19$_{-0.19}^{+0.10}$  &$81_{-4}^{+6}$  &$4.0_{-2.3}^{+1.9}$  &26$\pm8$       &1.82$\pm0.05$  &4.2$\pm0.5$         &4.9  &4     \smallskip\\
 42182 &42879  &SMC  &0.16$_{-0.16}^{+0.11}$  &$82_{-4}^{+5}$  &$3.4\pm1.7$        &24$_{-6}^{+7}$  &1.82$\pm0.05$  &4.1$\pm0.5$         &4.9  &4    \smallskip\\
 42182 &42879  &MW   &0.39 fixed          &$79\pm3$       & $7.4\pm0.9$        &$39\pm4$       &$1.73\pm0.06$  &$4.4_{-0.4}^{+0.5}$  &8.7  &5      \smallskip\\
 46630 &53807  &MW   &$0.39\pm0.03$           &$78\pm2$       &$7.2\pm1.1$        &$38\pm3$       &$1.77\pm0.05$  &$4.4\pm0.4$         &13   &5      \smallskip\\
 46630 &53807  &LMC  &$0.38\pm0.03$           &$78\pm2$       &$7.0\pm1.0$        &$38\pm3$       &$1.76\pm0.06$  &$4.4\pm0.4$         &16   &5     \smallskip\\
 46630 &53807  &SMC  &$0.37\pm0.04$           &$79\pm2$       &$6.2\pm0.9$        &$35\pm3$       &$1.77\pm0.05$  &$4.4\pm0.4$         &18   &5      \\
 \hline
\end{tabular}
\tablefoot{The model consists of a blackbody plus a power law, both
  attenuated by optical/NIR reddening and X-ray metal absorption.}
%\end{onehalfspace}
\end{minipage}
\end{table*}

%% file: c4_multi.tex
\section{Multicolour evolution of SN 2010bh}\label{sCurv}

We now present the entire GROND data set, which includes data in seven
different bands covering the wavelength range from 380 to 2300
nm. Photometry was corrected for host-galaxy extinction computed at
the end of Sect.\ \ref{ssExt}. In addition, the afterglow component
derived in Sect.\ \ref{ssMod} was subtracted from the data.  Assuming
a power-law decay for the optical afterglow with $\alpha=-1.3\pm0.2$
computed from the contemporaneous X-ray data, a flux and its
corresponding uncertainty were derived from the power-law model fitted
to the early SEDs for each epoch. We note however that typically
$\alpha_X\neq\alpha_{\rm opt}$. The correction for the afterglow
contribution is more significant at earlier times and in redder bands
and it has a lower significance at around maximum brightness for all
bands.

\subsection{Optical and near-infrared light curves}

\begin{figure}
  \centering
  \includegraphics[bb=22 154 457 714,width=\linewidth,clip]{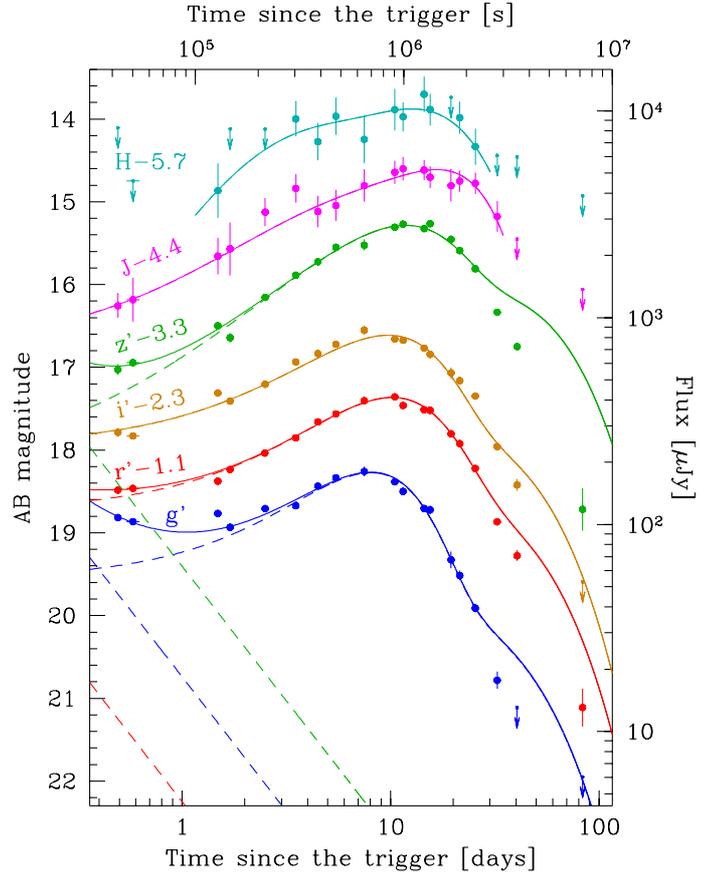}
  \caption{Multi-colour light curves corrected for host-galaxy
    extinction and afterglow-subtracted. Filled circles represent
    detections and arrows are upper limits. Solid lines represent the
    overall fits and dashed lines individual components. Fits of SN
    1998bw templates are extrapolated from $\approx t_0+60$~d. For
    reasons of clarity, light curves were shifted along the magnitude
    axis and the systematical error in \Ahost{V}\ was not added to the
    error bars.}
  \label{fLC}
\end{figure}

Figure \ref{fLC} shows the light curves in the optical \griz\ and the
NIR $JH$ bandpasses. The \grizJ\ light curves show the usual pattern
of SE SNe: the redder the filter, the later and broader the peak. The
$H$ band peaks a few days earlier than $J$, although given the large
uncertainties, the time difference is insignificant. The \K\ band
shows no credible detections in any of our observations down to limits
in the range of 18.4\rang18.9 mag (AB system) despite our observations
covering the expected peak of the SN (10.5\rang25.5~d after the
burst). In the following section, we analyse our data of SN 2010bh,
one of the best-observed GRB-SNe to date, and compare the light curve
with those of other SNe connected to GRBs.

\subsection{Colour evolution}

\begin{figure}
  \centering
  \includegraphics[bb=22 154 457 714,width=\linewidth,clip]{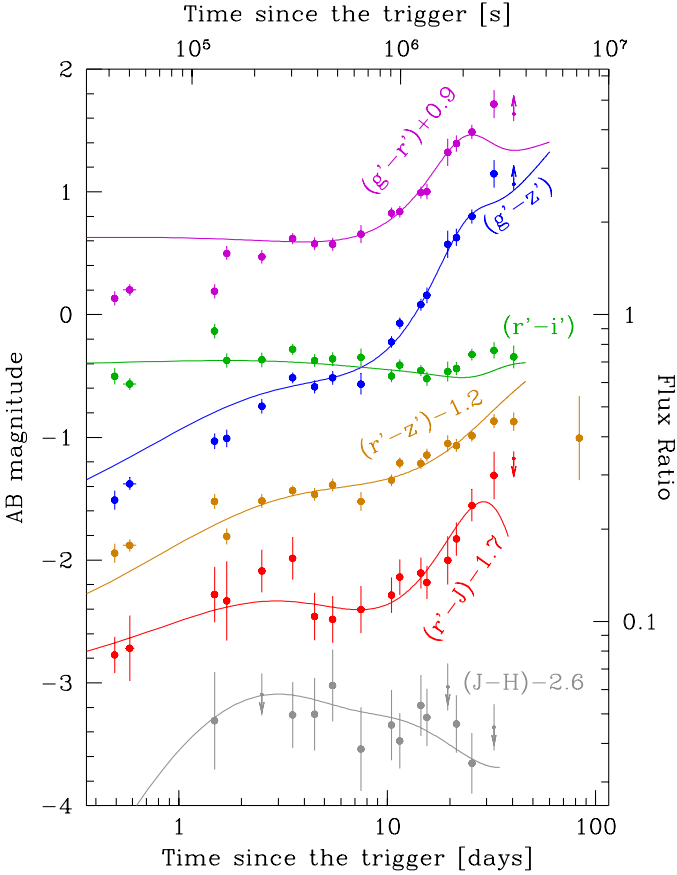}
  \caption{Colour curves corrected for host-galaxy extinction and
    afterglow-subtracted. Filled circles represent detections in both
    filters and arrows are upper limits derived from detections in one
    of the two bands. Solid lines represent the colour evolution of SN
    1998bw derived from the light-curve fits in Sect.\
    \ref{ss98bw}. For clarity, error bars do not include the
    systematic error in \Ahost{V} and for the last epoch, upper limits
    are not shown.}
  \label{fCC}
\end{figure}

Six GROND colour curves are presented in Fig.\ \ref{fCC} to analyse
the colour evolution of SN 2010bh. Most colours show drastic
evolution, where the cooling of the SN photosphere is evident as
colours become redder. Non-significant variations over time are shown
in \rmi\ and \JmH\ colours.

\subsection{Comparison with SN 1998bw}\label{ss98bw}

\input{t2_98bw}

To study the luminosity evolution of SN 2010bh to other SNe, we fit SN
1998bw templates derived from the publicly available $UVBRI$
photometry \citep{Galama98b} to each of our GROND filters. The process
of compiling these templates from the observed light curve of SN
1998bw is based on \citet{ZehKloHar04}. Given their modus operandi,
the NIR templates are inaccurate because they rely on an extrapolation
of the $UVBRI$ data. Therefore, the three epochs of \JHK\ data in
\citet{Patat01} were used to define the zero points of the flux
scale. For all filters, it is assumed that the host-galaxy extinction
of SN 1998bw is equal to zero \citep{Patat01,Clocchiatti11}. An
analytical function was then employed to parametrise the SN 1998bw
templates. Two of the seven template parameters represent the
brightness and morphology of the light curve, the factors of
luminosity $k$ and stretch $s$, respectively. These were defined
provided that $k=1$ and $s=1$ for SN 1998bw
\citep[][]{ZehKloHar04}. In addition, it was necessary to include a
delay parameter $t_{\rm delay}$ that acts by shifting the template
linearly in time, i.e., as a time offset. The delay parameters that
differ from zero do not necessarily mean that the onset of the SN is
out of phase.

Given that the early blue emission is inconsistent with the templates
even after subtraction of the afterglow component derived in Sect.\
\ref{ssMod}, an empirical power-law component was required (dashed in
Fig.\ \ref{fLC}). The power-law slope was fixed to
$\alpha=-1.3$. Leaving $\alpha$ free did not improve the fit and
different values of $\alpha$ only negligibly affected the stretch and
luminosity factors derived for the SN component. Table \ref{t98bw}
summarises the results of fitting the SN 1998bw templates and the
empirical power-law to the GROND data. The overall fits are shown in
Fig.\ \ref{fLC} for each band using solid lines.

Luminosity factors listed in the first line of Table \ref{t98bw},
which include the uncertainty in \Ahost{V}, reflect differences
between the colours of SN 2010bh and those of SN 1998bw. The
\prima{g}\ band is as bright as SN 1998bw and, as a rough comparison,
disagrees with the fainter $B$-band results from \citet{Cano11a} at
the $5\sigma$ confidence level after including the uncertainty in
their extinction determination. Our \ri\ luminosity factors are also
larger than previously reported from optical photometry of SN 2010bh
\citep[$k\approx0.4\rang0.5$;][]{Cano11a} mainly owing to the use of a
different extinction correction. There is certainly some intrinsic
blue excess in the \prima{g}\ band, although the difference from the
\prima{r} band of 40\% has a rather large uncertainty of 13\%. Higher
luminosities at bluer wavelengths indicate that the photosphere has a
higher temperature at its peak luminosity than that of SN
1998bw. Given that the expanding atmospheres of SNe cool down over
time and SN 2010bh peaks about a week earlier than SN 1998bw in
\prima{g}, it is a natural conclusion that it must be hotter at
maximum brightness. It is also possible that a \prima{g}-band peak as
early as 8~d after the trigger still contains a non-negligible
contribution from the cooling shock breakout. We also note that the
\prima{z}-band luminosity factor is much larger than in the other
bands; it is brighter than SN 1998bw at the $3\sigma$ confidence
level. After inspection of the spectra around maximum light presented
by \citet{Bufano11}, the \ion{Ca}{ii} $\lambda8579$ emission line was
found to contribute roughly 2\rang4\% to the continuum flux integrated
in the \prima{z}-band sensitivity range (8254\rang9528 \r A). The
contribution of this spectral feature is not enough to account for the
excess of 45\% in the \prima{z}\ band relative to the infrared. The
discrepancy is therefore attributed to the systematical uncertainties
introduced when extrapolating the $UBVRI$ data from \citet{Galama98b}
to construct the templates of SN 1998bw.  In summary, the luminosity
of SN 2010bh is a factor 0.5\rang0.7 fainter than SN 1998bw in optical
\ri\ bands and 0.6\rang0.7 in NIR $JH$ bands. We also note that at
late times SN 2010bh fades more rapidly in the optical than SN 1998bw
did, although the fluxes if the templates were extrapolated after
$\approx t_0+60$~d.

The stretch factors listed in the second line of Table \ref{t98bw}
range from 0.6 to 0.8 and are at the low end of the GRB-SNe
distribution \citep[see figure 5 of][]{Ferrero06}. These are also
consistent with the findings by \citet{Cano11a} for SN 2010bh. The
optical light curves are wider than the stretch factor suggests, or in
other words, the stretch factor predicts later peak times for the
optical light curves. To account for this, $t_{\rm delay}$ shifts the
templates to earlier times by about three days in \gri, which means
that the optical light curves of SN 2010bh peak even earlier than the
stretch factor suggests. After the inclusion of the delay parameter
$t_{\rm delay}$, the stretch factors now solely represent the width of
the light curves and not the time of the maximum brightness compared
to SN 1998bw. Moreover, peak times of roughly 8\rang9~d in the \gr\
bands (equivalent to the $V$ in the Johnson filter system) corresponds
to the earliest and fastest light curves of GRB-SNe observed to date
\citep[see table 2 of][]{Richardson09}.

Offsets in peak time, in our case represented by $t_{\rm delay}$, can
be explained as the result of delayed black-hole formation
\citep{VieSte99}. In this scenario, the XRF might be triggered by the
core collapse of the progenitor to a neutron star, soon after which
accretion holds. The supernova would then occur after the further
collapse of the neutron star into a black hole. The delay could be of
the order of months or years or perhaps as short as hours
\citep[see][]{ZehKloHar04}. Nevertheless, the time delay here is
negative and it is much more plausible that the comparison of
light-curve morphology to SN 1998bw may need more than two parameters
to be accurate.

Fits of SN 1998bw templates were also employed to study the colour
evolution of SN 2010bh in detail as presented in Fig.\ \ref{fCC},
where the curves are shown without the empirical afterglow
component. From the first two data points, it is clear from the \gmr\
and \gmz\ colours that there is a blue component that cannot be
modelled by the templates and is interpreted as the shock breakout in
Sect.\ \ref{ssMod}. From these two colours, it is also possible to see
that SN 2010bh becomes red faster than SN 1998bw did. The \rmi\ and
\JmH\ colours remain roughly constant, which shows that the changes
occur on a broader wavelength scale. The standard colour evolution
from blue to red is shown by \rmz\ and \rmJ, which at late times
evolve bluer and redder than SN 1998bw templates, respectively.

\subsection{Comparison with SN 2006aj}

We now compare results from optical data by \citet{Ferrero06}, who
used the same technique and templates to get the luminosity and
stretch factors for SN 2006aj. They computed luminosity factors in the
range 0.62\rang0.76, which are approximately in the same range as SN
2010bh without considering the measurements in \gz. Their stretch
factors ranged from about 0.62 to 0.69, which makes the SN 2010bh
optical light curves wider than those of SN 2006aj. In contrast to the
definition of the stretch factor, namely that earlier peak times tend
to correspond to a narrower light curve, peak times are earlier in the
case of SN 2010bh. Whilst the $BV$ photometry for SN 2006aj peaked
roughly 9 and 11~d after the burst, respectively, the \gr\ photometry
for SN 2010bh peaks approximately 8 and 9~d after trigger,
respectively. This supports the statement that SN 2010bh has evolved
more rapidly than any other GRB-SNe, given that SN 2006aj had the
fastest evolution until the discovery of SN 2010bh. The $JH$ light
curves of SN 2006aj \citep{Cobb06,Kocevski07} are scaled to the
luminosity distance of SN 2010bh for comparison. After the
host-extinction correction, SN 2010bh turns out to be as bright as SN
2006aj in the NIR as well. Hence, SN 2010bh is similar to SN 2006aj in
terms of light-curve shape and luminosity.

%% file: t2_98bw.tex
\begin{table*}
\begin{minipage}[t]{\textwidth}
  \caption{Fits of SN 1998bw templates to SN 2010bh.}\label{t98bw}
  \centering
%\begin{onehalfspace}
\begin{tabular}{l c c c c c c c}
  \hline\hline\noalign{\vspace{0.5\smallskipamount}}
  & \prima{g}  & \prima{r}  & \prima{i}  & \prima{z} & $J$  & $H$  & \K \vspace{0.5\smallskipamount}  \\
  \hline\noalign{\vspace{0.5\smallskipamount}}
  SN amplitude ($k$\,)\,\footnote{For afterglow-subtracted and host-extinction-corrected data ($\Ahost{V}=1.2\pm0.1$ mag).} &$1.10\pm0.12$ &$0.65\pm0.05$ &$0.54\pm0.04$ &$1.25\pm0.07$ &$0.69\pm0.07$ &$0.63\pm0.08$  &$<1.4$ \vspace{0.4\smallskipamount}  \\
% SN amplitude ($k$\,)\,\tablefootmark{a} &$1.10\pm0.12$ &$0.65\pm0.05$ &$0.54\pm0.04$ &$1.25\pm0.07$ &$0.69\pm0.07$ &$0.63\pm0.08$  &$<1.4$   \\
  Stretch factor ($s$\,) &$0.77\pm0.01$ &$0.78\pm0.01$ &$0.65\pm0.02$ &$0.73\pm0.02$ &$0.84\pm0.04$ &$0.67\pm0.07$ &\nodata   \\
  Peak delay ($t_{\rm delay}$)\,\footnote{Relative to maximum luminosity of SN 1998bw after applying a time offset due to the stretch factor.} &$-3.3$ &$-3.2$ &$-2.6$ &$-0.4$ &$-1.0$ &0 &\nodata  \vspace{0.5\smallskipamount} \\
% Peak delay ($t_{\rm delay}$)\,\tablefootmark{b} &$-3.3$ &$-3.2$ &$-2.6$ &$-0.4$ &$-1.0$ &0 &\nodata   \\
  \hline
\end{tabular}
%\tablefoot{
%  \tablefoottext{a}{For afterglow-subtracted and
%    host-extinction-corrected data ($\Ahost{V}=1.2\pm0.1$ mag).}
%  \tablefoottext{b}{Relative to maximum luminosity of SN 1998bw after
%    applying time offset due to the stretch factor.}
%}
%\end{onehalfspace}
%\vspace{-2mm}
\end{minipage}
\end{table*}

%% file: c5_bolo.tex
\section{Bolometric light curve}\label{sBol}

The bolometric light curves of SNe are an essential tool for examining
global luminosity features and enable us to compare with other SNe and
theoretical models. However, it is difficult to obtain such a light
curve because of the limited information at UV, infrared, and radio
wavelengths. Only a \emph{quasi}-bolometric light curve can be
constructed by using a broad spectral coverage to derive a total flux
that is then used as a proxy of the bolometric flux. To accomplish
this task, we employed the wavelength range covered by our \grizJH\
filters, i.e., from 350 to 1800 nm.

In the case of SN 2010bh, there are no UV constraints. In the
following description, no attempts at correcting for the UV flux were
made. By using the afterglow-subtracted \grizJH\ photometry corrected
for host-galaxy extinction, monochromatic fluxes for each bandpass
were derived. Sets of three bandpasses were defined to interpolate
their corresponding monochromatic fluxes using the Simpson's rule. The
second-degree polynomial result of the interpolation was then
integrated over frequency in the range of each set of
bandpasses. Finally, the total flux in the range from 350 to 1800 nm
was determined by adding up the integrated fluxes of all sets of
bandpasses. The total flux is transformed to a quasi-bolometric
luminosity using a distance of $240\pm17$ Mpc to the host galaxy of SN
2010bh (see Sect.\ \ref{ssMod} for more details). No attempts of
extrapolation beyond the limits of the \prima{g}\ and the $H$
bandpasses were made. Corrections for the NIR flux at late times were
found to be the most significant. The data at $t_0+30.7$~d (rest
frame) were corrected for the $H$-band non-detection by assuming that
the fraction of $H$-band flux compared to the total bolometric flux
remains constant at 8\% starting from $t_0+24.1$~d. Similar
corrections were performed for $JH$ non-detections at 38.6 and 78.8~d
after the burst under the assumption of a constant $JH$ flux fraction
of 27\% at $t_0+30.7$~d. Results are shown in Fig.\ \ref{fBol} along
with quasi-bolometric light curves from SE and other GRB-SNe.

\input{fBol1} %%sidecaption

The analysis of the quasi-bolometric light-curve morphology yields a
peak luminosity of $4.3\times$10$^{42}$ \eps\ at about 8~d after the
trigger (equivalent to $M_{bol}\approx-17.87$), i.e., approximately
two times fainter and six days sooner than for SN 1998bw. Our
luminosity is 16\% higher than that computed by \citet{Cano11a},
although consistent to within our 11\% of uncertainty. Whilst the
early peak of SN 2010bh correlates with its narrowness and low
luminosity, this is the case for neither the entire GRB-SNe sample nor
the local sample of SE SNe
\citep{ZehKloHar04,RicBraBar06,Richardson09}. The morphology of the
light curve is similar to that of SN 2006aj \citep{Pian06}, although
21\% fainter. The peak time also resembles that of the type-Ic SN
1994I \citep{Richmond96}, although SN 2010bh has a much wider light
curve, which is 77\% brighter at maximum. In terms of peak luminosity,
SN 2010bh is similar to the broad-lined Ic SN 2009bb
\citep{Pignata11}. It also underwent the most dramatic late-time decay
in the sample, which implies that its envelope became rapidly
optically thin to $\gamma$-rays. The last statement is supported by
the extremely high expansion velocities measured for SN 2010bh of the
order of 30,000 \kps\ \citep{Chornock10c}. Another clear feature is
the sudden decrease in luminosity at around $t_0+30$~d, which
contrasts with the smooth decay in the comparison SNe at similar
stages. This indicates either that the atmosphere becomes rapidly
optically thin or that the assumption of a constant NIR contribution
after $\approx t_0+31$~d underestimates the flux in the $JH$ bands.

\subsection{Physical parameters of the explosion}\label{ssPhy}
We followed the approach described in \citet{Valenti08} to derive the
physical quantities that characterise the explosion, i.e., we modelled
the early and late light curves separately. The early-time phase
corresponds to the photospheric regime for which the analytical model
developed by \citet{Arnett82} has been adopted, initially used for SNe
Ia and adapted to SE SNe
\citep[e.g.,][]{Taubenberger06,Valenti08,Pignata11,Benetti11}. At late
stages, the atmosphere becomes nebular, i.e., optically thin, and the
emitted luminosity is powered by the energy deposition of: (1)
$\gamma$-rays from $^{56}$Co decay, (2) $\gamma$-rays from
electron-positron annihilation, and (3) the kinetic energy of the
positrons \citep[see appendix A in][]{Valenti08}. However,
\citet{Maeda03} noted that the two-component configuration leads to
inconsistencies between the parameters derived from fitting the early
and late light curves of SNe Ic, caused mainly by the model
limitations in varying the $\gamma$-ray trapping over time. To enable
low and high $\gamma$-ray trapping at early and late times,
respectively, \citet{Maeda03} divided the ejecta into a high-density
inner region and a low-density outer region. The emission from the
outer region dominates the total emission in the optically thick
regime at early times, and that from the inner region, which has a
higher $\gamma$-ray opacity, dominates in the nebular phase at late
times. Here, we use the same procedure to model the
\grizJH\ quasi-bolometric light curve of SN 2010bh.

Given the model explained above, a total of four free parameters were
used to fit the quasi-bolometric light curve of SN 2010bh: the total
mass of $^{56}$Ni produced in the envelope $M_{\rm Ni}$, the total
ejecta mass $M_{\rm ej}$, the fraction of mass in the inner component
$f_M$, and the fraction of kinetic energy in the inner component
$f_E$. The kinetic-energy-to-ejected-mass ratio of the outer region
was fixed by using its correlation with photospheric velocity at peak
luminosity \citep{Arnett82}
\begin{align}
  \upsilon_{\rm ph}^2&\approx\frac{3}{5}\,\frac{2E_{\rm k,out}}{M_{\rm
      ej,out}}.
  \label{eVph}
\end{align}
\noindent This expression assumes that the density of the ejecta is
homogeneous and that the inner component does not contribute to the
emitted luminosity in the optically-thick regime. Since the
photospheric velocity was not available directly from observations,
the velocity measured by fitting P-Cygni line profiles was used as a
proxy of $\upsilon_{\rm ph}$. However, the envelope layer where the
blue-shifted absorption line forms does not necessarily coincide with
the position of the photosphere, as found when measuring different
expansion velocities from absorption lines of different species. The
spread can amount to several hundreds \kps\ (see \citealt{Jones09} for
an example of type-II SNe). For the spectra of SN 2010bh,
\citet{Chornock10c} obtained velocities of about 35,000 and 26,000
\kps\ from the \ion{Si}{ii} $\lambda6355$ feature roughly 21 and 6~d
after the burst, respectively. Since there is no measurement at the
time of maximum light, which is about 8\rang9~d after the trigger, a
range of expansion velocities was used in the modelling. Assuming that
the photosphere lies at deeper layers than those where lines are
formed and recedes in mass exposing deeper and slower layers, in the
modelling we employed photospheric expansion velocities of 2.5, 2.8,
and 3.1 $\times10^4$ \kps. Other physical and mathematical quantities
such as opacity and integration constants were chosen to be the same
as in \citet{Cano11a}.

\input{t3_phy}

Our fitting procedure consisted of two steps. First of all, we
modelled the data around maximum luminosity ($5<t-t_0\leq30$~d) using
the Arnett's model, assuming that only the outer component contributes
to the total luminosity at this stage. We then obtained $M_{\rm
  Ni,out}$ and $M_{\rm ej,out}$. Secondly, we modelled the late-time
data ($t-t_0>30$) using the nebular-phase components of
\citet{Valenti08} assuming that both the inner and outer regions
contribute to the total emitted luminosity. Here, we fixed $M_{\rm
  Ni,out}$ and $M_{\rm ej,out}$ to the values obtained in the first
step of the fitting procedure and only $f_M$ and $f_E$ were allowed to
vary.

From the above modelling scheme, our best-fit parameters were $M_{\rm
  Ni,out}=(0.135\pm0.001)M_{\odot}$, $M_{\rm ej,out}=2.37\rang2.90$
$M_{\odot}$, $f_M=0.36\pm0.04$, and $f_E=0.11\rang0.15$ for the three
selected expansion velocities at the photosphere. All of these results
combined together provided the total masses and energy of the
explosion listed in Table \ref{tPhy} (statistical errors only). The
$^{56}$Ni mass is independent of the chosen expansion velocity at the
photosphere, $M_{\rm Ni}=(0.21\pm0.03)M_{\odot}$, given that it is
proportional to the luminosity. In contrast, since the ejected mass
changes significantly, the weighted mean and the RMS of the results of
the three models in Table \ref{tPhy} were employed to compute a final
value of $M_{\rm ej}=(2.60\pm0.23)M_{\odot}$. The total kinetic energy
and energy fraction were derived in the same way, implying that
$E_{\rm k}=(2.4\pm0.7)\times10^{52}$ erg and $f_E=0.12\pm0.02$.

Using comparable independent data, \citet{Cano11a} followed a similar
procedure to derive physical quantities from quasi-bolometric data.
Whilst our values for $M_{\rm ej}$ are consistent with those in
\citet{Cano11a} to within $1.5\sigma$, they found that $M_{\rm
  Ni}=(0.10\pm0.01)M_{\odot}$ for SN 2010bh, which is two times lower
than our value. The discrepancy affects the determination of the
kinetic energy as well, which is connected in Eq.\ \ref{eVph} to
$\upsilon_{\rm ph}$ and $M_{\rm ej}$. The causes of these
inconsistencies are: (1) the host-galaxy extinction employed, which is
$\EBVhost=0.39\pm0.03$ mag in our case and $0.18\pm0.08$ mag in
\citet{Cano11a}; (2) the inclusion of an inner component hidden at
optically thick stages, which increases the $^{56}$Ni mass by a factor
of $f_M$; and (3) the choice of different expansion velocities of the
photosphere. Furthermore, even when including the uncertainties in the
host-extinction determination, which are $\sigma_{M_{\rm Ni}}\sim0.02$
both here and for \citet{Cano11a}, the discrepancy
persisted. Nevertheless, evidence for the existence of a dense inner
layer of $0.94\pm0.15 M_{\odot}$ is provided and supported by the
sub-luminous post-maximum phase of SN 2010bh, which is indicative of a
high trapping of $\gamma$-rays.

The large amount of $^{56}$Ni produced is consistent with the class of
SN associated with GRBs, but also matches the value derived for the
highly energetic type-Ic SN 2004aw \citep{Valenti08}, which in
contrast had a much broader light curve that varied more slowly with
time. Most interestingly, SN 2010bh resembles SN 2006aj in terms of
light-curve shape. The $^{56}$Ni mass produced in the explosion is
practically the same \citep{Mazzali06b}, whilst the ejected mass is
20\% higher for SN 2010bh and the total kinetic energy is
significantly different ($\sim2\times10^{51}$ erg in the case of SN
2006aj). This ensures that SN 2010bh is remarkable in terms of
expansion velocity, which has been one of the greatest ever measured
\citep{Chornock10c}.

We note that the end of the optically-thick regime and the beginning
of the nebular phase cannot be accurately defined. The rule of thumb
is that the photospheric phase ends 30~d after explosion at the
earliest and that the nebular phase starts 60~d after the explosion at
the latest. Since Arnett's model fits our data relatively well until
day 30, it is defined as the end of the optically-thick
era. Nevertheless, we are unable to clearly establish whether the
period between 30 and 60~d after the burst corresponds to the nebular
phase already, as the ejecta are most probably neither completely
thick nor sufficiently thin.

The model reproduces the luminosity at maximum brightness and
thereafter, despite underpredicting the luminosity during the rising
phase. This is because the light curve is dominated by shock-breakout
emission at early stages. The same is obtained in Sect.\ \ref{sCurv},
when trying to fit templates of SN 1998bw to our multicolour light
curve, where an extra component for the shock breakout is
required. The light curves of SN 2010bh peak so early that an even
more rapid evolution than observed is expected after maximum
luminosity. This early-wide peak dichotomy could be explained by the
$^{56}$Ni distribution in the envelope. If there were more $^{56}$Ni
produced near the surface, the peak would be early and wide, i.e., the
SN would rise more slowly but much earlier. In contrast, if the
$^{56}$Ni were concentrated towards the centre, it would lead to a
much later peak, although sharper rise, producing a narrower shape of
the maximum (M.~Bersten, 2010, private communication;
\citealt{Nomoto10}).

%% file: fBol1.tex
\begin{figure*}
  \sidecaption
  \includegraphics[bb=19 276 591 710,width=0.70\linewidth,clip]{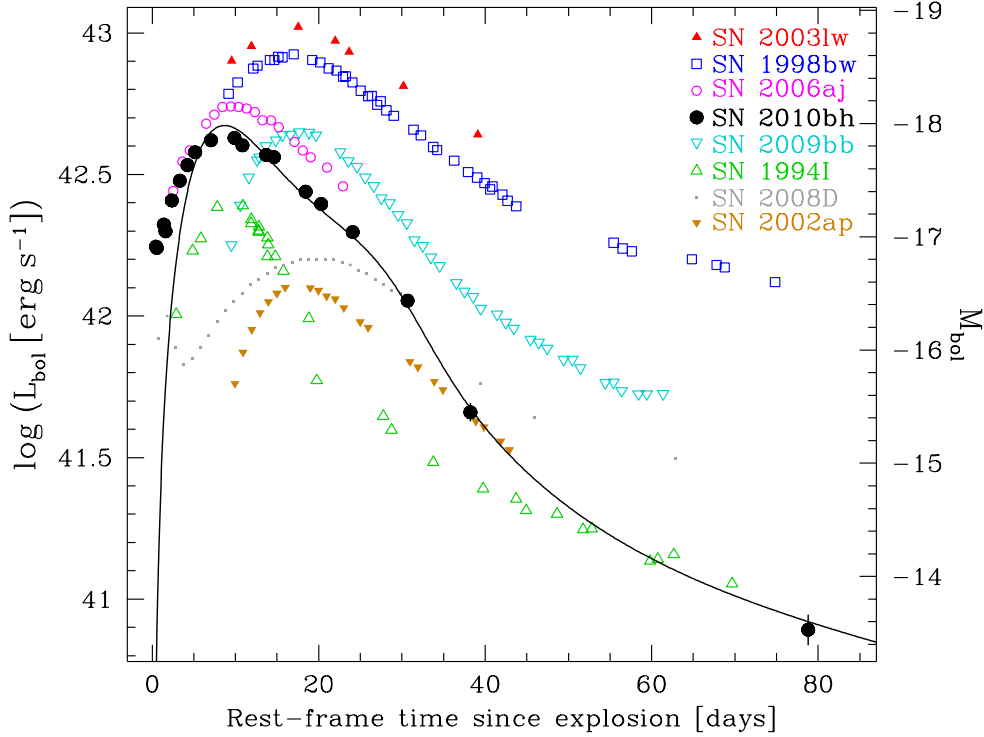}
  \caption{Quasi-bolometric light curve of SN 2010bh produced by using
    GROND \grizJH\ filters (black open circles) in the rest frame. For
    clarity, the uncertainties in the host-galaxy extinction are not
    included in the error bars. A single black continuous line
    represents the best-fit model (see Sect.\ \ref{ssPhy}). Early and
    late components of the model are smoothly joined at $t_0+30$~d. SE
    and other GRB-SNe quasi-bolometric light curves have been plotted
    as a comparison sample: SN 2003lw \citep[GRB
    031203;][]{Malesani04}, SN 1998bw \citep[GRB
    980425;][]{Galama98b}, SN 2006aj \citep[GRB 060218;][]{Pian06},
    the broad-lined Ic SN 2009bb \citep{Pignata11}, the type-Ic SN
    1994I \citep{Richmond96}, the type Ibc SN 2008D \citep[XRO
    080109;][]{Modjaz09,Soderberg08}, and the type-Ic SN 2002ap
    \citep{GalOfeShe02,Foley03,Yoshii03}.}
  \label{fBol}
\end{figure*}

%% Alternative to sidecaption

%\begin{figure*}
%\begin{center}
%  \begin{minipage}[t]{0.65\linewidth}
%    \includegraphics[bb=19 154 590 586,width=\linewidth,clip]{bol.eps}
%  \end{minipage}\hfill
%  \begin{minipage}[b]{0.33\linewidth}
%    \caption{Quasi-bolometric light curve of SN 2010bh produced by
%      using GROND \grizJH\ filters (black open circles) in the rest
%      frame. For clarity the uncertainties in the host-galaxy
%      extinction are not included in the error bars. A single black
%      continuous line represents the best-fit model (see Sect.\
%      \ref{ssPhy}). Early and late components of the model are
%      smoothly joined at $t_0+30$~d. SE and other GRB-SNe
%      quasi-bolometric light curves have been plotted as comparison
%      sample: SN 2003lw \citep[GRB 031203;][]{Malesani04}, SN 1998bw
%      \citep[GRB 980425;][]{Galama98b}, SN 2006aj \citep[GRB
%      060218;][]{Pian06}, the broad-lined Ic SN 2009bb
%      \citep{Pignata11}, the type-Ic SN 1994I \citep{Richmond96}, the
%      type Ibc SN 2008D \citep[XRO 080109;][]{Soderberg08}, and the
%      type-Ic SN 2002ap
%      \citep{GalOfeShe02,Foley03,Yoshii03}.\label{fBol}}
%  \end{minipage}
%\end{center}
%\end{figure*}

%% file: t3_phy.tex
\begin{table*}
\begin{minipage}[t]{\textwidth}
  \caption{Physical parameters of the explosion with a varying
    expansion velocity.}\label{tPhy} \centering
%\begin{onehalfspace}
\begin{tabular}{c c c c c c c c}
\hline\hline\noalign{\vspace{0.5\smallskipamount}}
  $\upsilon_{\rm ph}$   &$M_{\rm Ni,out}$   &$M_{\rm ej,out}$   &$f_M$          &$f_E$          &$M_{\rm Ni}$   &$M_{\rm ej}$   &$E_{\rm  k}$    \\
  $[{\rm km\ s}^{-1}]$  &[$M_\odot$]    &[$M_\odot$]            &               &               &[$M_\odot$]    &[$M_\odot$]    &[$10^{52}$ erg] \vspace{0.4\smallskipamount}\\
\hline\noalign{\vspace{0.4\smallskipamount}}
  25,000                &$0.135\pm0.001$  &$1.52\pm0.05$        &$0.36\pm0.02$  &$0.15\pm0.03$  &$0.21\pm0.02$  &$2.37\pm0.10$  &$1.85\pm0.09$    \\
  28,000                &$0.135\pm0.001$  &$1.70\pm0.05$        &$0.36\pm0.04$  &$0.13\pm0.07$  &$0.21\pm0.03$  &$2.64\pm0.14$  &$2.52\pm0.22$    \\
  31,000                &$0.135\pm0.001$  &$1.87\pm0.06$        &$0.36\pm0.02$  &$0.11\pm0.02$  &$0.21\pm0.02$  &$2.90\pm0.12$  &$3.37\pm0.13$    \\
\hline
\end{tabular}
%\end{onehalfspace}
%\vspace{-2mm}
\end{minipage}
\end{table*}

%% file: c6_concl.tex
\section{Conclusions}\label{sCon}

Spanning a time range from 12 hours to 83 days after the trigger and
covering from 190 to 2300 nm in wavelength (see Sect.\ \ref{sObs}), we
have presented UV/optical/NIR photometric data of XRF 100316D/SN
2010bh. Given the results introduced and discussed in Sect.\
\ref{sSED}, \ref{sCurv}, and \ref{sBol}, we have drawn the following
conclusions:

\begin{itemize}
\item Broad-band SEDs at early times demonstrate the existence of red
  and blue components identified as synchrotron emission from the XRF
  afterglow and the cooling envelope after shock breakout,
  respectively.
\item A significant amount of dust along the line of sight through the
  host galaxy ($\Ahost{V}=1.2\pm0.1$ mag) is consistent with the
  two-component SED model and agrees with the faint detections at UV
  wavelengths.
\item By comparing with earlier X-ray results from \citet{Starling11},
  we have demonstrated that the temperature of the blackbody component
  decreases with time, which is consistent with a scenario of a
  cooling expanding atmosphere.
\item By performing an additional analysis of the thermal component
  and the earlier X-ray measurements we have measured expansion
  velocities that are consistent with SN expansion and an initial
  apparent emission radius of $7\times10^{11}$ cm. This radius is
  slightly larger than the size of WR stars, which are the most likely
  GRB-SNe progenitors. If a WR star were the progenitor of XRF
  100316D/SN 2010bh, then the initial radius could indicate that there
  was a massive dense stellar wind surrounding the progenitor.
\item Our multicolour light curves after host-galaxy correction and
  subtraction of the afterglow component, have peak \ri\ luminosities
  of about 0.5\rang0.7 times that of SN 1998bw and consistent with
  those of SN 2006aj. Similarly the NIR luminosity at maximum is as
  bright as SN 2006aj and 0.6\rang0.7 times that of SN 1998bw. The
  excess in the \prima{g}\ band indicates that SN 2010bh has a hotter
  photosphere than that of SN 1998bw at the time of maximum
  brightness.
\item We have found that SN 2010bh is the most rapidly evolving
  GRB-SNe to date, reaching maximum luminosity 8\rang9 days after the
  burst in the \gr\ bands.  At late times, it also fades more rapidly
  than SN 1998bw showing redder colours as well. This behaviour is
  also evident in the bolometric light curve, which decays faster than
  for any SN in the comparison sample.
\item The physical parameters of the explosion are derived by means of
  the quasi-bolometric light curve constructed from our \grizJH\
  photometry. The modelling is performed using Arnett's model
  \citep{Arnett82} for data around peak and standard $\gamma$-ray
  deposition at later times. A high-density inner component with
  roughly 26\% of the total mass is required to reproduce the flux
  ratio between maximum luminosity and tail. The total mass of
  $^{56}$Ni produced in the envelope is $M_{\rm Ni}=0.21\pm0.03
  M_\odot$, which precisely matches the value derived for SN 2006aj,
  whilst the total ejecta mass of $M_{\rm ej}=2.6\pm0.2 M_\odot$
  exceeds the value for SN 2006aj by 20\%. However, the kinetic energy
  turns out to be higher at $E_k=(2.4\pm0.7)\times10^{52}$ erg, making
  SN 2010bh the second most energetic GRB-SN after SN1998bw.
\end{itemize}

The association between XRF 100316D and SN 2010bh is particularly
interesting, since for the second time the cooling of the shock
breakout has been detected in a GRB-SN. It is also unique in revealing
a hot component that possibly contributes even at blue maximum
brightness, one of the largest host-galaxy extinctions measured for
this kind of transient, and the fastest rise among GRB-connected SNe.

%% file: zAck.tex
\begin{acknowledgements}

  F.~O.~E. thanks Ferdinando Patat for stimulating discussion as well
  as for accurate comments and suggestions. We thank the referee and
  editors for accurate comments and useful suggestions. Part of the
  funding for GROND (both hardware and personnel) was generously
  granted from the Leibniz-Prize to Prof.~G.\ Hasinger,
  \emph{Deut\-sche For\-schungs\-ge\-mein\-schaft} (DFG) grant HA
  1850/28--1. This work made use of data supplied by the UK
  \emph{Swift} Science Data Centre at the University of Leicester. The
  Ph.~D.\ studies of F.~O.~E.\ are funded by the German
  \emph{Deut\-scher Aka\-de\-mi\-scher Aus\-tausch Dienst}, DAAD, and
  the Chilean \emph{Co\-mi\-si\'on Na\-cion\-al de
    In\-ves\-ti\-ga\-ci\'on Cien\-t\'{\i}\-fi\-ca y
    Tec\-no\-l\'o\-gi\-ca}, CONICYT. T.~K.\ acknowledges support by
  the DFG cluster of excellence ``Origin and Structure of the
  Universe'' and by the European Commission under the Marie Curie
  Intra-European Fellowship Programme.  The Dark Cosmology Centre is
  funded by the Danish National Research Foundation.  S.~K., A.~N.~G.,
  A.~Rossi, and D.~A.~K.\ acknowledge support by DFG grant Kl
  766/16--1. M.~N.\ acknowledges support by DFG grant SA 2001/2--1.
  P.~S.\ acknowledges support by DFG grant SA 2001/1--1.  A.~C.~U.,
  A.~N.~G., D.~A.~K., and A.~Rossi are grateful for travel funding
  support through MPE. Figure \ref{fFC} is partially based on
  observations made with the NASA/ESA Hubble Space Telescope, obtained
  from the data archive at the Space Telescope Science
  Institute. STScI is operated by the Association of Universities for
  Research in Astronomy, Inc.~under NASA contract NAS 5--26555. This
  research has made use of NASA's Astrophysics Data System.

\end{acknowledgements}

%% file: zA1_data.tex
\begin{appendix}

\section{Optical and near-infrared photometry}

\begin{table}[h!]
\begin{minipage}[t]{18cm}
  \caption{GROND photometry of field stars used for relative
    photometry.}\label{tStd}
  \centering
%\begin{onehalfspace}
\begin{tabular}{ccccccccc}
 \hline\hline\noalign{\vspace{0.5\smallskipamount}}
 R.A.(J2000) &Dec.(J2000) &\prima{g}  &\prima{r}  &\prima{i}  &\prima{z} &$J$  &$H$ &\K  \\
 $[\ ^\mathrm{h}\ $:$\ ^\mathrm{m}\ $:$\ ^\mathrm{s}$ ] &[ $^\circ$ : $'$ : $''$ ]  & & & & & & & \\
 \hline\noalign{\vspace{0.4\smallskipamount}}
 07:10:30.10 &$-56$:14:58.1 &17.981(16) &17.423(10) &17.246(12) &17.065(16) &16.587(12) &16.737(13) &\nodata  \\
 07:10:34.37 &$-56$:15:58.1 &18.092(16) &17.365(10) &17.158(11) &16.973(16) &16.477(12) &16.646(13) &16.964(31) \\
 07:10:28.77 &$-56$:14:30.9 &16.221(13) &15.831(10) &15.742(11) &15.596(15) &15.224(07) &15.404(08) &15.933(15) \\
 07:10:31.17 &$-56$:14:45.8 &16.232(12) &15.535(09) &15.305(10) &15.094(15) &14.606(06) &14.642(07) &15.083(11) \\
 07:10:29.97 &$-56$:16:24.4 &17.703(13) &16.435(10) &15.902(10) &15.570(15) &14.875(06) &14.864(07) &15.216(11) \\
 07:10:30.98 &$-56$:14:26.3 &16.804(13) &16.391(10) &16.267(11) &16.117(16) &15.733(09) &15.949(09) &16.374(22) \\
 \hline
\end{tabular}
\tablefoot{All magnitudes are in the AB system and corrected for
  neither Galactic nor host-galaxy extinction. Errors are statistical
  only.}
%\end{onehalfspace}
%\vspace{-2mm}
\end{minipage}
\end{table}

\begin{table}[h!]
\begin{minipage}[t]{18cm}
  \caption{GROND photometry of XRF 100316D/SN 2010bh after image
    subtraction.}\label{tData}
  \centering
%\begin{onehalfspace}
\begin{tabular}{r@{\,--\,}lcccccc}
 \hline\hline\noalign{\vspace{0.5\smallskipamount}}
 \multicolumn{2}{c}{Time Interval}  &$g\,'$  &$r\,'$  &$i\,'$  &$z\,'$ &$J$  &$H$  \\
 \multicolumn{2}{c}{[d] after the trigger} & & & & & & \\
 \hline\noalign{\vspace{0.4\smallskipamount}}
0.49023  &0.49470  	&20.75(04)	&20.90(04)	&20.87(05)	&20.77(07)	&20.69(14)	&$>20.04$ \\
0.56046  &0.60240  	&20.81(03)	&20.91(03)	&20.94(04)	&20.76(04)	&20.71(26)	&$>20.67$ \\
1.47634  &1.49123  	&20.75(04)	&20.90(04)	&20.59(05)	&20.51(05)	&20.47(22)	&20.69(33) \\
1.69356  &1.70396  	&20.92(04)	&20.77(03)	&20.69(05)	&20.65(06)	&20.40(32)	&$>20.09$ \\
2.49804  &2.51041  	&20.70(04)	&20.58(04)	&20.50(05)	&20.20(04)	&19.99(17)	&$>20.09$ \\
3.50557  &3.52164  	&20.67(03)	&20.40(03)	&20.24(03)	&19.94(03)	&19.72(17)	&19.94(21) \\
4.48027  &4.48963  	&20.43(04)	&20.21(03)	&20.15(04)	&19.78(04)	&20.00(19)	&20.22(23) \\
5.46324  &5.47229  	&20.33(04)	&20.11(04)	&20.04(04)	&19.61(04)	&19.93(19)	&19.92(23) \\
7.47679  &7.48174  	&20.26(06)	&19.96(04)	&19.87(06)	&19.58(06)	&19.69(19)	&20.20(28) \\
10.47338 &10.48259 	&20.38(03)	&19.91(03)	&19.98(04)	&19.37(03)	&19.53(14)	&19.85(25) \\
11.47550 &11.47994 	&20.50(03)	&20.01(03)	&19.99(04)	&19.33(03)	&19.49(14)	&19.94(18) \\
14.47912 &14.49638 	&20.71(04)	&20.07(03)	&20.08(03)	&19.38(03)	&19.51(12)	&19.67(22) \\
15.48101 &15.48997 	&20.72(05)	&20.07(04)	&20.16(05)	&19.32(04)	&19.59(13)	&19.85(19) \\
19.52885 &19.53123 	&21.33(10)	&20.36(05)	&20.39(07)	&19.51(05)	&19.70(19)	&$>19.71$ \\
21.48630 &21.49543 	&21.52(06)	&20.48(04)	&20.48(04)	&19.65(03)	&19.64(13)	&19.95(19) \\
25.49771 &25.51874 	&21.91(05)	&20.78(03)	&20.66(03)	&19.87(03)	&19.67(13)	&20.30(21) \\
32.48587 &32.50697 	&22.78(10)	&21.42(04)	&21.28(05)	&20.39(04)	&20.07(19)	&$>20.41$ \\
40.46892 &40.50042 	&$>23.11$	&21.83(06)	&21.74(07)	&20.81(05)	&$>20.34$	&$>20.43$ \\
83.43058 &83.45161 	&$>23.95$	&23.66(23)	&$>22.91$	&22.78(25)	&$>20.95$	&$>20.90$ \\
 \hline
\end{tabular}
\tablefoot{The template image for subtraction was taken 234 days after
  burst. The time reference is $t_0=55271.53113$ MJD, the date of the
  GRB trigger. All magnitudes are in the AB system and corrected for
  neither Galactic nor host-galaxy extinction. Upper limits are all at
  the 3$\sigma$ confidence level. Errors include the systematics of
  the photometric calibration.}
%\end{onehalfspace}
%\vspace{-2mm}
\end{minipage}
\end{table}

\end{appendix}